\documentclass[12pt,twoside,openright,a4paper]{report}
\usepackage{epsf}
\usepackage[dvips]{epsfig}
\usepackage{amssymb}
\usepackage{german}

\newcommand{\fn}[1]{\begin{equation} #1 \end{equation}}

\newcommand{\subs}[1]{_{\mbox{\scriptsize #1}}}
\newcommand{\sups}[1]{^{\mbox{\scriptsize #1}}}
\newcommand{\bbar}[1]{\overline{#1}}
\newcommand{\ha}{hadronic axion }
\newcommand{\has}{hadronic axions }
\newcommand{\ad}{accretion disc }
\newcommand{\ads}{accretion discs }

\newcommand{\ns}{neutron star }
\newcommand{\nss}{neutron stars }

\newcommand{\simgt}{\,\rlap{\lower 5.5 pt \hbox{$\mathchar \sim$}} \raise 
	1pt \hbox {$>$}\,}
\newcommand{\simlt}{\,\rlap{\lower 5.5 pt \hbox{$\mathchar \sim$}} \raise
	1pt \hbox {$<$}\,}
\begin{document}
\originalTeX
\renewcommand {\theequation}{\arabic{chapter}.\arabic{equation}}
\pagenumbering{roman}
\begin{titlepage}
\begin{center}
\vspace{3cm}
{\huge \normalsize\huge Impact of Hadronic Axions
on Black~Hole\\[0.3cm] Accretion Discs and Neutron Stars}\\
\vspace{3cm}
{\Large Diplomarbeit}\\
{\large von}\\
{\Large Klaus B\"ocker}\\
\vspace{3cm}
{\large M\"unchen, September 1999}\\
\vspace{3.5cm}
{\unitlength1mm
\begin{picture}(70,35)
\put(20,0){\psfig{file=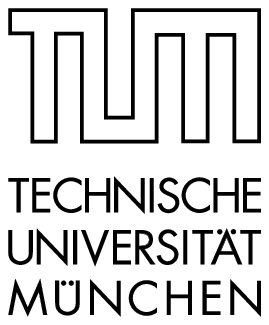,width=3.0cm}}
\end{picture}}\\
\vspace{1cm}
{\normalsize Physik-Department der Technischen Universit\"at M\"unchen\\
Institut f\"ur Theoretische Physik T30\\
Prof.\ Dr.\ Manfred Lindner\\
Angefertigt am Max-Planck-Institut f\"ur Physik\\
(Werner-Heisenberg-Institut)\\
bei Dr.\ Georg Raffelt}
\newpage \thispagestyle{empty} \rule{0cm}{1cm}
\end{center}
\end{titlepage}\newpage
\germanTeX
\begin{center}
{\bf Zusammenfassung}
\end{center}
\vspace{0.8ex}
{\small Axionen entstehen bei der L\"osung des "`starken
CP-Problems"' mit Hilfe der Peccei-Quinn-Symmetrie.  
Der astrophysikalisch und kosmologisch prinzipiell erlaubte
Bereich von Axionenmassen beschr\"ankt sich auf zwei Regionen:
zum einen "`leichte"' Axionen
mit Massen zwischen $10^{-5}$ und
$10^{-3}$ eV, die kalte dunkle Materie (CDM) des Universums
darstellen, zum anderen
"`schwere"' Axionen aus dem 
sogenannten "`hadronischen Axio\-nen\-fen\-ster"' mit Massen
um 10 eV.
Ihre Existenz w\"urde mit $\Omega_a \approx$ 0.1--0.3 zur
hei\ss en dunklen Ma\-te\-rie (HDM) beitragen.

Die genaue Breite des hadronischen Axionenfensters h\"angt stark vom
zugrunde\-lie\-gen\-den Axionenmodell ab.  Deswegen werden in dieser Arbeit
zun\"achst die be\-ste\-hen\-den und f\"ur hadronische Axionen relevanten
astrophysikalischen Argumente neu \"uber\-pr\"uft. Das Ergebnis ist, dass
Axionen mit einer Masse zwischen 10 und 20 eV nicht aus\-ge\-schlos\-sen
werden k\"onnen, falls sie
keine Baum-Niveau-Kopplungen an
gew\"ohn\-liche Quarks und Leptonen besitzen und wenn 
gleichzeitig ihre
Kopplung an Photonen ext\-rem stark unterdr\"uckt ist.

Hadronische Axionen wechselwirken in erster Linie mit Protonen und Neutronen.
Als Konsequenz k\"onnen sie in nukleonenreicher Materie durch den
Brems\-strah\-lungs\-pro\-zess $N+ N \rightarrow N+ N+ a$ erzeugt werden. 
Daher -- um 
das hadronische Axionenfenster genauer zu untersuchen -- betrachten 
wir Akkretionsscheiben um Schwarze L\"ocher 
und isolierte Neutronensterne.
Erstere k\"onnen w\"ahrend des Kollaps eines Neutronen\-dop\-pel\-sterns
entstehen und liefern ein vielversprechendes Modell
f\"ur kurze Gam\-ma\-strah\-lungs\-aus\-br\"uche. Gem\"a\ss\ diesem entstehen
in der Akkretionsscheibe Neutrinos und Antineutrinos, von 
denen sich ein Teil in Elektron-Positron-Paare und weiter in hoch\-en\-er\-ge\-ti\-sche 
Gam\-ma\-strah\-lung umwandelt. Ein zus\"atzlicher Ener\-gie\-ver\-lust durch die
Emission hadronischer Axionen k\"onnte die Luminosit\"at der 
Gammastrahlungsausbr\"uche erheblich verringern. Wir stel\-len fest, dass
Axionen mit einer Masse von etwa 11 bis 66~eV einen wichtigen 
Einfluss auf Gammastrahlungsausbr\"uche haben, 
wobei diese quantitativen Aussagen jedoch stark
vom zugrundeliegenden Modell der Ak\-kre\-tions\-scheibe und dem des Schwarzen Lochs
abh\"angen.

Signifikantere Grenzen f\"ur die Masse hadronischer Axionen ergeben sich aus
der Betrachtung
isolierter Neutronensterne. Mit einem einfachen Modell
f\"ur das
Ab\-k\"uh\-lungs\-ver\-hal\-ten eines Neutronensterns zeigen wir, 
dass die gemessenen
Ober\-fl\"a\-chen\-tem\-pe\-ra\-turen der Pulsare 
PSR 1055-52, PSR 0630+178 (Gemina) und PSR 0656+14 zu hoch sind, um
mit der Existenz hadronischer Axionen im Einklang zu stehen. Diese
Er\-kennt\-nis wird durch numerische Berechnungen der Neutronensternk\"uhlung
von Umeda et al.\ \cite{UmedoIwamoto} best\"atigt, deren Ergebnisse
wir einer erweiterten Interpretation im Hinblick auf hadronische
Axionen unterziehen.}
\originalTeX
\begin{center}
{\bf Acknowledgment}
\end{center}
\vspace{0.8ex}
I am greatly indebted to my supervisor, Georg Raffelt, 
for introducing me
to this intriguing subject, and for his support and advice.

Furthermore, I have benefited from the friendly and stimulating atmosphere
at the Max-Planck-Institut f\"ur Physik, M\"unchen. Thanks to
all the students and postdocs, especially
to Robert Buras for stimulating discussions, and Maurice
Lausberg and Bj\"orn P\"otter for teaching me to play table tennis.

Last but not least, I would like to thank my partner Katharina Macketanz
for putting up with me.

\tableofcontents
\newpage

\pagenumbering{arabic}
\setcounter{equation}{0}  \chapter{Introduction}
The axion was proposed in 1977 to solve the CP problem of QCD (``strong
CP problem''). 
Since then, more than 500 research papers about axions and related topics
have been published \cite{SpiresSearch}. 
Evidently, axions have 
been a stimulating subject!
The main reason for this steady interest is that axions not only 
provide an
elegant solution to the strong CP problem, but are also attractive 
from a cosmological perspective. More specifically, there are two
regions of axion masses where these particles 
could constitute some or all of the cosmic
dark matter: Axions with a mass in the range  
$m_a \approx 10^{-5}$--$10^{-3}\mbox{ eV}$ are an ideal candidate 
for cold dark matter (CDM), and
two current experiments actually search for these very 
light particles \cite{CDMAxionExpA,CDMAxionExpLLNL}. 
Moreover, it was recently claimed that ``heavy'' axions 
in the so-called ``hadronic axion window'' around
$10\mbox{ eV}$
provide a hot dark matter (HDM) component with $\Omega_a \approx 0.1$--0.3,
exactly the amount needed in mixed dark matter scenarios
\cite{Moroietal}. 

While there is no doubt that axions in
the CDM window are astrophysically and cosmologically
allowed \cite{RaffeltPPfS}, the situation is less clear for 
the HDM range because the existence and exact width of the
hadronic axion gap strongly depends on details of the axion model. 
Therefore, firstly
we have re-examined the hadronic axion window and have found
that axions without tree-level couplings
to ordinary quarks and leptons (``KSVZ-type axions'') and masses between
10 and 20\mbox{ eV} are not excluded by previous astrophysical arguments 
if their coupling to photons 
$g_{a\gamma\gamma}$ is less than approximately $3 \times 10^{-11}
\mbox{ GeV}^{-1}$. This value is
nearly one hundred times smaller
than in generic axion models, implying a strong
suppression of the photon coupling for these models
to be viable in the hadronic mass range. 
It turns out that one can construct models where this suppression arises as
a accidental cancellation
between two independent contributions to the axion-photon 
coupling \cite{KaplanHA}.
Therefore the hadronic axion window is indeed
open. Henceforth, the term ``hadronic axions'' shall refer to models
with severely suppressed photon couplings, 
i.e.\ they are a subclass of ``KSVZ-type axions,'' a term which
applies to models without tree-level coupling to ordinary quarks and leptons.

Hadronic axions are very weakly interacting particles, and 
one might think that they can not have any significant 
influence on astrophysical
objects. However,
as a consequence of the axion-pion mixing, their effective 
nucleon coupling is comparable to those of generic axion models.
This means that those astrophysical objects
which mainly consist of nuclear matter are suitable candidates to 
explore the hadronic axion window.
Therefore, we consider isolated neutron stars 
and black hole accretion discs (BHADs) as axion laboratories---both 
types of objects have not been studied in this context.

Accretion discs around black holes are
possible outcomes of neutron star mergers and have recently been 
discussed 
in connection with a very intriguing class of 
mysterious astrophysical phenomena:
the gamma-ray bursts, short and intense photon eruptions
with energies of typically 100 keV--1 MeV\@.
According to the BHAD model, a gamma-ray burst develops when neutrinos, which
are produced inside the hot accretion torus, are converted
into a photon burst.
The crucial point is that the initial neutrino luminosity must be 
extraordinarily large
to account 
for the observed energies in photons in the range $10^{50}$--$10^{53}
\mbox{ erg}$. However, numerical
neutron star merger simulations \cite{JankaNEU}
yield total neutrino luminosities of about $10^{53}\mbox{ erg s}^{-1}$, enough
to explain short and weak gamma-ray bursts with energies in the ballpark 
of $10^{50}$--$10^{51}\mbox{ erg}$.
This situation could change if axions existed in the \ha window. 
Produced via the
bremsstrahlung process $N + N \rightarrow N + N + a$, they
represent a novel, important energy-loss mechanism in accretion discs.
In this case, the neutrino luminosity would be significantly 
reduced, dimming the observed gamma-ray burst signal.
Put another way, 
the requirement that the axion luminosity must not exceed the 
neutrino emission 
puts an upper axion mass limit, which we find in 
the range 10--100\mbox{ eV}, i.e. \has could have an impact
on the evolution of gamma-ray bursts. However, this
bound depends on unknown details of the BHAD model. Nevertheless, although
precise and reliable statements about the influence of \has 
on gamma-ray bursts require a deeper understanding of these fascinating 
celestial objects, it is quite certain that \has if they were found
have to be included in present BHAD models.

Finally, we discuss the cooling of neutron stars by means of 
hadronic axion emission. We set up a simple cooling model
and show that recently measured surface temperatures of
the isolated neutron stars
PSR 1055-52, PSR 0630+178 (Gemina), and PSR 0656+14 are 
too high to be in accordance with hadronic axion emission because
these particles significantly accelerate the energy loss of the star.
Furthermore, we review a recent calculation of
Umeda et al.\ \cite{UmedoIwamoto} who studied neutron star cooling
including axions by means of numerical simulations. However, they 
did not discuss the implications for hadronic axions.
It turns out that an interpretation of their results with
regard to the hadronic axion window
confirms the outcome of our simple model, namely 
that hadronic axion emission is in conflict with
the observed surface temperatures
of neutron stars.

\setcounter{equation}{0} \chapter{Summary of Axion Physics\label{chap1}}
By way of introduction, we briefly summarize the most important 
aspects of axion physics. First, 
we outline how axions solve the strong CP problem, and 
how they arise as a simple extension of the Standard Model. Then,
the KSVZ axion is discussed in more detail.
Finally, the couplings 
of axions to ordinary matter, i.e.\ to photons, electrons and nucleons
are discussed.
\section[Motivation for Axions: The Strong CP
Problem]{Motivation for Axions: The Strong\\ CP Problem \label{ABHOA}}
In weak interactions one observes CP 
violation, whereas the strong interactions 
are known to respect P and CP invariance to very high accuracy.
This CP invariance of the 
strong sector is a problem
because QCD predicts CP violation 
as a consequence of two different mechanisms. First, there is the topologically
nontrivial ground state of
QCD, the so-called ``$\Theta$-vacuum''. Second, the axial
transformation, necessary to diagonalize the complex quark mass matrix, also
leads to CP violation.
Both effects together cause an 
additional nonperturbative term in the QCD Lagrangian,
\fn{{\cal L}\subs{SM}\sups{eff}= {\cal L}\subs{SM}\sups{pert} + \bar{\Theta} 
\frac{g_s^2}{32 \pi^2}\, 
G_a^{\mu \nu} \,\tilde{G}_{\mu \nu}^a\,,\label{Lageins}} 
where
${\cal L}\subs{SM}\sups{pert}$ is the perturbative part of the Standard 
Model Lagrangian, $g_s$ the strong 
coupling constant, $G_a^{\mu \nu}$ the color field strength tensor, 
$\tilde{G}^a_{\mu \nu}\equiv \frac{1}{2}\epsilon_{\mu 
\nu \rho \sigma}\,G^{a\rho \sigma}$ its dual,
and $\bar{\Theta}=\Theta + \mbox{Arg det} M$.
The $\epsilon$ tensor that occurs in the definition of 
$\tilde{G}$ implies that ${\cal L}\subs{SM}\sups{eff}$ is not 
invariant under parity, and thus odd under CP.
Note that the effective CP violating parameter $\bar{\Theta}$ is the sum of 
a QCD contribution---\mbox{the vacuum 
angle $\Theta$}---and an electroweak part---\mbox{Arg det $M$}---related 
to the 
phase structure of the quark mass 
matrix $M$. 
The experimental bound on the neutron electric dipole moment
requires that
\fn{|\Theta + \mbox{Arg det} M| \simlt 10^{-9}.}
Now,
it is difficult to understand why a compound quantity like $\bar{\Theta}$, 
which is a sum of two very different 
contributions, should be so small. This fine-tuning problem
is known as the strong CP 
problem.

The strong CP problem can be elegantly explained by introducing an additional 
global, chiral symmetry with an 
associated current and charge: the Peccei-Quinn-symmetry 
$U(1)\subs{PQ}$ and the Peccei-Quinn 
charge \cite{PQ}. Its existence requires certain 
enhancements of the Higgs sector of the 
Standard Model so that its enlarged 
symmetry group becomes $SU(3)_C \times SU(2)_L \times U(1)_Y \times U(1)
\subs{PQ}$. However, it should be stressed that $U(1)\subs{PQ}$ is a global 
symmetry 
and not a gauge symmetry.
In typical axion models, 
PQ symmetry is achieved by introducing additional Higgs fields $\Phi$ with
degenerate vacua; at 
least one is 
necessary.
However, in the real world the chiral symmetry $U(1)\subs{PQ}$ is not 
observed, hence it must be 
spontaneously broken at an energy scale $f_a$; the associated 
Nambu-Goldstone boson is the axion. Under a PQ 
transformation the axion field $a(x)$ shifts as
\fn{a(x) \rightarrow a(x) + \alpha\, f_a\, ,\label{aft}}
where $\alpha$ is the parameter associated with the $U(1)\subs{PQ}$ 
transformation.
Now we come to the essential feature of the PQ mechanism: the PQ symmetry 
$U(1)\subs{PQ}$ is only exact at the classical level. At the
quantum level, however, it is broken by the 
Adler-Bell-Jackiw anomaly 
because  the PQ symmetry is
chiral. As a result, an additional term appears 
in the effective Lagrangian of equation (\ref{Lageins}),
\fn{{\cal L}\subs{SM}\sups{eff} = {\cal L}\subs{SM}\sups{pert} + \bar{\Theta} 
\frac{g_s^2}{32 \pi^2}\,
G_a^{\mu \nu} \,\tilde{G}^a_{\mu \nu} 
+
 \xi\, \frac{a}{f_a} \frac{g^2}{32 \pi^2}\,G_a^{\mu \nu} \,\tilde{G}^a_{\mu 
\nu}\, ,\label{lagrangian}}
where $\xi$ is a model dependent parameter.
The $\xi$-term is very similar to the CP violating $\bar{\Theta}$-term and 
provides an effective potential $V\subs{eff}(a)$ 
for the axion field $a(x)$. This potential implies 
that the degeneracy of the vacuum is removed. 
Peccei and Quinn showed that the vacuum expectation 
value of the axion field, i.e.\ the 
value of $a(x)$
where $V\subs{eff}(a)$ has its minimum, is given by
\fn{\langle a(x)\rangle = - \frac{f_a}{\xi} \bar{\Theta}\,.\label{vacneu}}
Now one introduces a physical axion field $a\subs{phy}$, which is defined 
as excitations around the vacuum expectation 
value (\ref{vacneu}), i.e.\
\fn{a(x)\subs{phy} \equiv a(x) - \langle a(x) \rangle\,,}
and substitutes it into expression (\ref{lagrangian}). 
In doing this, the 
$\bar{\Theta} G \tilde{G}$ term is eliminated, and therefore the strong 
CP problem is solved. Furthermore, one is left with an axion-gluon 
interaction of the form $a\subs{phy}\, G \tilde{G}$---obviously an 
immediate outcome of every axion model.
As a result of this generic axion-gluon coupling, axions pick
up a small mass of order
\begin{eqnarray}
m_a& \approx &\frac{m_\pi f_\pi}{f_a}\nonumber\\
& \approx &6 \mbox{ eV} \,\left(\frac{10^6 \mbox{ GeV}}{f_a}\right), 
\label{axionmassformula}\end{eqnarray}
where $f_\pi \approx$ 93 MeV is the pion decay constant and 
$m_\pi = 135\mbox{ MeV}$ the pion mass. 
As the axion has a mass, it is actually a ``pseudo 
Nambu-Goldstone boson.'' 

In summary, we arrive at the following Lagrangian, where
from now on $a$ stands for the physical axion field $a\subs{phy}$,
\begin{eqnarray}
{\cal L}\sups{eff}\subs{SM}& = &{\cal L}\subs{SM}  
+ \frac{1}{2} \partial_\mu  a\, \partial^\mu a \nonumber\\
& &+ {\cal L}\subs{int}\left(\frac{\partial_\mu a}{f_a},\, 
\psi\right) +\xi \frac{a}{f_a} 
\frac{g^2}{32 \pi^2}\,G_a^{\mu \nu} \,\tilde{G}^a_{\mu \nu}\ .\label{finallag}
\end{eqnarray}
The second term is the kinetic energy of the axion field, the third 
represents possible 
interactions of the axion with fermions $\psi$ and 
has to be in agreement with the 
classical invariance under $U(1)
\subs{PQ}$, i.e.\ under the transformation (\ref{aft}). 
Hence we conclude that
the axion field $a$ may enter the interaction Lagrangian 
${\cal L}\subs{int}$ only
through derivative terms $\partial_\mu a$.
\section{Axion Models}
\subsection{Standard Axion and Invisible Axions}
Let us now look at different axion models which are based on these general 
ideas. 
First, we consider the question how the axial symmetry 
$U(1)\subs{PQ}$, which solves the strong
CP problem,
can be reconciled with the Standard Model. 
In the Standard Model one introduces a complex scalar field, the 
well-known Higgs field $\phi$. It is an $SU(2)_L$ doublet, i.e.\ its weak 
isospin is $I_W = \frac{1}{2}$. Furthermore,
its weak hypercharge is $Y=1$. As a consequence of the Higgs mechanism, all 
boson and fermion masses are 
generated by this Higgs 
field $\phi$. The four phases of the isodoublet $\phi$ provide four 
additional degrees of freedom. Three of them 
provide the longitudinal degrees of the $W^\pm$ and $Z^0$, and the remaining 
phase ends up as the Higgs boson. 
So, in the Standard Model there is no room for axions. Another way to see  
this is to recall the 
transformation properties of the Standard-Model Lagrangian under axial 
transformations. In particular, consider 
the Yukawa-coupling terms
\fn{{\cal L}\subs{Yuk} = -G_d\, \bar{q}_L\, \phi\, d_R - G_u\, \bar{q}_L\, 
\phi_c\, u_R + \mbox{h.c.}\ ,
\label{yukawa}}
where $\bar{q}=(\bar{u},\bar{d})$ and 
$\phi_c \equiv i \tau_2 \phi^\ast$. One can ensure that the first term is 
invariant under the special axial 
$U(1)\subs{PQ} = U(1)_R - U(1)_L$ transformation
\footnote{Consider a general
axial transformation $U(1)_A$ and its 
related axial vector 
current $j^\mu_A = \bar{\psi}\gamma^\mu\gamma^5 
\psi$. With $\psi_L \equiv 
\frac{1}{2}(1-\gamma^5)\psi$ and $\psi_R \equiv \frac{1}{2}(1+\gamma^5)\psi$ 
one can write the axial current in 
terms of left- and right-handed currents $j^\mu_L$ and $j^\mu_R$, $$j^\mu_A = 
(\psi_R + \psi_L)\gamma^\mu
\gamma^5(\psi_R + \psi_L)=
 \bar{\psi}_R\gamma^\mu \psi_R - \bar{\psi}_L\gamma^\mu \psi_L = j^\mu_R - 
j^\mu_L,$$
where we have used $\gamma^5 \psi_R=\psi_R,\, \gamma^5 \psi_L=-\psi_L$
and $\bar{\psi}_L\gamma^\mu\psi_R = \bar{\psi}_R\gamma^\mu\psi_L =0$.
Therefore we may write $U(1)_A = U(1)_R - U(1)_L$.} 
$q_L \rightarrow e^{i \alpha /2}\,q_L$, $d_R 
\rightarrow e^{-i \alpha /2}
\,d_R$,
$u_R \rightarrow e^{-i \alpha /2}\,u_R$
if one requires the Higgs field $\phi$ to transform as
\fn{\phi \rightarrow e^{i \alpha}\,\phi .\label{higgstrans}}
However, the second term in (\ref{yukawa}) is only invariant if $\phi_c$ 
transforms in the same way as $\phi$, 
but this is not the case. 

A simple solution is to introduce a second Higgs doublet $\phi_2$ in place 
of $\phi_c$, which is 
therefore independent of $\phi_1 \equiv \phi$. Then, both Higgs fields 
$\phi_1$ and 
$\phi_2$ 
transform as in (\ref{higgstrans}), and PQ invariance of (\ref{yukawa}) 
is 
attained. Finally, when the  $SU(2)_L \times U(1)_Y \times U(1)\subs{PQ}
$ symmetry 
breaks spontaneously, the 
massless 
axion appears
together with three other Nambu-Goldstone bosons, which will be ``eaten''  
by the massive  $W^\pm$ and $Z^0$ gauge 
bosons. This is the original standard-axion model in which the breaking scale of 
the PQ symmetry 
is equal to the electroweak one, 
i.e.\ $f_a \approx 250 \mbox{ GeV}$. The associated axion with a mass of about 
$100 \mbox{ keV}$, however, was quickly ruled 
out by experiments: if weak-scale axions were to exist, one would expect 
decays in the meson system like
$$K^\pm \rightarrow \pi^\pm + a\ .$$
The absence of such decays implies that the standard axion can not exist. 
Even worse, its mass 
must be less than about 
10 keV, or equivalently the symmetry-breaking scale must be larger than 1000 
GeV. 

In order to maintain the PQ solution of the strong CP problem, 
the ``invisible axion'' was invented. 
The special feature 
of these models is that the axion resides in the phase of a new complex 
$SU(2)_L \times U(1)_Y$ singlet scalar 
field $\Phi$, which has a nontrivial PQ charge
$Q_\Phi$. As a consequence, the field $\Phi$ does not participate in weak 
interactions, and hence 
the PQ symmetry breaking is decoupled from the electroweak one. Therefore, 
the PQ scale $f_a$ 
is an arbitrary
parameter in invisible axion models,
implying also that the axion's couplings are not fixed. 
Perhaps the most common axion models are those of Dine, Fischler, Srednicki, 
and Zhitnitski\u{\i} (DFSZ) and Kim, 
Shifman, Vainshtein, and Zakharov (KSVZ). The KSVZ axion is an example of a 
wider class of axion models, 
the so-called ``KSVZ-type axions.'' The main difference 
between the DSVZ and KSVZ-type models is that the latter have no tree-level 
couplings to ordinary quarks and leptons.
This is achieved by setting the PQ charge of all ordinary fermions to zero. 
Instead, one introduces new, heavy, colored
fermions with a nonvanishing PQ charge $X_f$.
\subsection{The KSVZ Model\label{KSVZgenau}}
\begin{figure}[tb]
\unitlength1mm
\begin{picture}(70,30)
\put(33,-25){\psfig{file=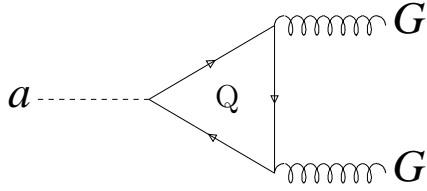,width=8.0cm}}
\put(72,14){{\cal Q}}
\end{picture}\par
\caption{Interaction of axions with gluons via a heavy quark loop.
\label{axiongluon}}
\end{figure}

In order to gain deeper insight into axion physics, one specific 
example
shall be discussed in some detail, i.e.\ the KSVZ model.  
Since the KSVZ axion has no tree-level coupling to ordinary quarks,
the question of whether such an axion is able to 
solve the strong CP problem arises: axions need an anomalous triangle 
coupling to two gluons, 
i.e.\ a  $U(1)\subs{PQ}$-$SU(3)_C$-$SU(3)_C$ anomaly, which is a fundamental
feature of every axion model. 
Therefore, within the KSVZ model, one has to 
introduce at least one new 
heavy Quark ${\cal Q}$ with nonvanishing PQ charge and nontrivial
transformation properties under $SU(3)_C$,
e.g.\ a $SU(3)_C$ triplet. As a result, the essential $a G \tilde{G}$ 
coupling can be 
realized at lowest order via an anomalous heavy quark triangle loop, which 
is illustrated 
in Fig.\ \ref{axiongluon}.   

Let us now consider the spontaneous breaking of the $U(1)\subs{PQ}$ 
symmetry in the KSVZ model. 
To this end, we introduce a new field $\Phi$ with the associated Lagrangian
\begin{eqnarray}
{\cal L}_\Phi &=& (\partial_\mu \Phi)^\dagger (\partial^\mu \Phi) - V(\Phi)
\nonumber\\
 &=& (\partial_\mu \Phi)^\dagger (\partial^\mu \Phi) +f\subs{PQ}^2 
\Phi^\dagger \Phi -\lambda 
(\Phi^\dagger \Phi)^2,
\label{LagPhi}
\end{eqnarray} 
where $\lambda, f\subs{PQ}^2 >0$. Below we set $\lambda = 1$. 
${\cal L}_\Phi$ is invariant under 
\fn{\Phi\rightarrow e^{i \alpha} \Phi
\label{PQPhi},} i.e.\ 
${\cal L}_\Phi$ possesses a $U(1)$ global symmetry, the PQ symmetry 
$U(1)\subs{PQ}$.
The potential $V(\Phi)$ then has a circle of minima at 
$\langle \Phi \rangle = (f\subs{PQ}/\sqrt{2}) 
e^{i \varphi}$ with an arbitrary phase $\varphi$. Now we spontaneously 
break the $U(1)\subs{PQ}$ symmetry by 
expanding $\Phi$ around one of these ground states $\langle \Phi \rangle$ 
with minimum energy. 
If we introduce new fields
$a$ and $\rho$ corresponding to   
angular 
and radial excitations around $\langle \Phi \rangle$, we may write
\fn{\Phi = \frac{1}{\sqrt{2}}(f\subs{PQ} + \rho) \exp\left(\frac{i a}
{f\subs{PQ}}
\right) .\label{PhiVaxExp}}
The massless mode $a$ is 
linked with the axion field and
translates under $U(1)\subs{PQ}$ in (\ref{PQPhi}) as $a \rightarrow a + 
\alpha f\subs{PQ}$.
Now we substitute (\ref{PhiVaxExp}) into (\ref{LagPhi}), i.e.\ 
we express the Lagrangian ${\cal L}_\Phi$ in terms of 
$a$ and $\rho$. The result is
\begin{eqnarray}
{\cal L}_\Phi &=& \frac{1}{2}(\partial_\mu a)(\partial^\mu a) 
+ \frac{1}{2}(\partial_\mu \rho)(\partial^\mu \rho)  
- f\subs{PQ}^2 \rho^2
-f\subs{PQ} \rho^3 
-\frac{\rho^4}{4} \nonumber \\
&&+\frac{1}{f\subs{PQ}} \rho\, (\partial_\mu a)(\partial^\mu a)
+\frac{1}{2 f\subs{PQ}^2} \rho^2 (\partial_\mu a)(\partial^\mu a)
+\mbox{const.}\label{Laggi} 
\end{eqnarray}
The third term has the form of a mass term $(-\frac{1}{2} m_\rho^2 \rho^2)$ 
for the $\rho$ field. Thus, the
$\rho$ mass is $m_\rho = \sqrt{2}\,f\subs{PQ}$. Because $f\subs{PQ}$ is 
a high-energy scale, $\rho$ is a very 
heavy particle and can be neglected as far as the low-energy regime is 
concerned.  
Therefore, we may keep only the axionic degree of freedom, i.e.\ we drop 
all terms involving $\rho$.

Finally, we have to implement the quark ${\cal Q}$ into our model. To this 
end, we add
a usual Dirac term for the fermionic ${\cal Q}$ field and introduce Yukawa 
couplings between ${\cal Q}$ and 
$\Phi$. We then write
\fn{{\cal L}_{\cal Q} = \frac{i}{2}\left(\bar{{\cal Q}}\gamma^\mu 
\partial_\mu{\cal Q} + \mbox{h.c.}\right)
  - h(\bar{{\cal Q}}_L {\cal Q}_R \,\Phi +
 \Phi^\dagger \bar{{\cal Q}}_R {\cal Q}_L)\label{Yukawa}}  
with a Yukawa coupling constant $h>0$.
Remember, the $U(1)\subs{PQ}$ transformation of the $\Phi$ field is already 
fixed through (\ref{PQPhi}). Hence, 
the transformation 
property of the quark field ${\cal Q}$ is determined by the demand for 
$U(1)\subs{PQ}$ invariance of the 
Yukawa interactions (\ref{Yukawa}). This invariance can be achieved by
\fn{{\cal Q}_L\rightarrow e^{i \alpha/2}{\cal Q}_L,\qquad {\cal Q}_R
\rightarrow 
e^{-i \alpha/2}{\cal Q}_R , \label{QPQtrafo}}
which is equivalent to the axial transformation 
${\cal Q}_R + {\cal Q}_L \equiv 
{\cal Q}\rightarrow e^{-i \gamma^5 \alpha/2}{\cal Q}$. Note
that both ${\cal Q}$ and $\bar{{\cal Q}}$ transform
in the same way, i.e.\ it is also $\bar{{\cal Q}}
\rightarrow \bar{{\cal Q}}\,e^{-i 
\gamma^5 \alpha/2}$.
If we insert (\ref{PhiVaxExp}) into (\ref{Yukawa}), we obtain for the Yukawa 
term, i.e.\ the last term 
in (\ref{Yukawa}),
\fn{{\cal L}\subs{Yuk} = -\frac{f\subs{PQ}}{\sqrt{2}}\, h 
\left(\bar{{\cal Q}}_L {\cal Q}_R\, e^{i a/f\subs{PQ}}+
e^{-i a/f\subs{PQ}}\,\bar{{\cal Q}}_R {\cal Q}_L\right) ,}
where we have neglected all terms involving $\rho$. 
One can simplify this expression by using
 the normal field ${\cal Q}={\cal Q}_R + {\cal Q}_L$
instead of the chiral quark fields ${\cal Q}_L$ and
${\cal Q}_R$. Recalling that
\fn{\bar{{\cal Q}}_L {\cal Q}_R =\bar{{\cal Q}}\, \frac{1}{2}(1 +\gamma^5) 
{\cal Q}\quad \mbox{and}
\quad \bar{{\cal Q}}_R {\cal Q}_L=\bar{{\cal Q}}\, \frac{1}{2}(1 -\gamma^5) 
{\cal Q} ,}
one finds
\begin{eqnarray}
{\cal L}\subs{Yuk} &=&- \frac{f\subs{PQ}}{\sqrt{2}}\, h\, \bar{{\cal Q}}
\nonumber
\left[e^{i a/f\subs{PQ}}\,\frac{1 +\gamma^5}{2} +e^{-i a/f\subs{PQ}}\,
\frac{1 -\gamma^5}{2}\right] {\cal Q}\\
\nonumber
 &=&- \frac{f\subs{PQ}}{\sqrt{2}}\, h\,\bar{{\cal Q}}
\left[\frac{e^{i a/f\subs{PQ}} +e^{-i a/f\subs{PQ}}}{2} + \gamma^5 
\frac{e^{i a/f\subs{PQ}} -e^{-i a/f\subs{PQ}}}{2}\right]{\cal Q}\\\nonumber
&=&- \frac{f\subs{PQ}}{\sqrt{2}}\, h\, \bar{{\cal Q}}\left[
\cos(a/f\subs{PQ}) + i \gamma^5 \sin(a/f\subs{PQ})\right] {\cal Q}\\
&=&- \frac{f\subs{PQ}}{\sqrt{2}}\, h\, \bar{{\cal Q}}\, e^{i \gamma^5 a/f
\subs{PQ}}\, {\cal Q}\ .\label{YukTwo}
\end{eqnarray}
With (\ref{Laggi}), (\ref{Yukawa}), and (\ref{YukTwo}) we finally get the 
KSVZ Lagrangian 
\fn{{\cal L}\subs{KSVZ}=\frac{i}{2}\left(\bar{{\cal Q}}\gamma^\mu 
\partial_\mu{\cal Q} + \mbox{h.c.}\right)
 + \frac{1}{2}(\partial_\mu a)^2 - 
\frac{f\subs{PQ}}{\sqrt{2}}\, h\, \bar{{\cal Q}}\, e^{i \gamma^5 a/f\subs{PQ}} 
\,{\cal Q}\ .\label{KSVZLag}}  
In order to interpret the last term of this expression, we expand it in 
powers of $a/f\subs{PQ}$. The zeroth-order 
term provides a mass for the quark ${\cal Q}$. Higher-order terms describe 
interactions between ${\cal Q}$ and 
the axion $a$. One obtains
\fn{{\cal L}\subs{KSVZ}=\frac{i}{2}\left(\bar{{\cal Q}}\gamma^\mu 
\partial_\mu{\cal Q} + \mbox{h.c.}\right)+
 \frac{1}{2}(\partial_\mu a)^2 - 
\frac{f\subs{PQ}}{\sqrt{2}}\, h\,\bar{{\cal Q}}{\cal Q} + {\cal L}
\subs{int}\ ,}
and thus a ${\cal Q}$ mass of $m_{\cal Q}=h f\subs{PQ}/\sqrt{2}$. In 
addition, the interaction Lagrangian is given by
\fn{{\cal L}\subs{int}= -i\frac{m_{\cal Q}}{f\subs{PQ}} a \bar{{\cal Q}}
\gamma^5 {\cal Q} 
+ \frac{m_{\cal Q}}{2 f\subs{PQ}^2} a^2 \bar{{\cal Q}}{\cal Q}+\ldots .
\label{IntOne}}
One then can see that to lowest order the interaction between
the axion $a$ and the exotic quark ${\cal Q}$ is pseudoscalar. The 
corresponding Yukawa coupling constant is $g_{a{\cal Q}} \equiv m_{\cal Q}/ 
f\subs{PQ}$. There are, however, 
higher-order
terms, which sometimes must be taken into account in order to achieve 
the correct result.
Therefore, it is better to search for an alternative approach to the 
$a$-${\cal Q}$ interaction.
One may remove the axion field in the Yukawa
coupling term (last term in (\ref{KSVZLag})) by a local chiral transformation 
of the ${\cal Q}$ field,
\fn{ {\cal Q}\rightarrow e^{-i \gamma^5 a/ 2 f\subs{\tiny PQ}}\, {\cal Q},
\qquad \bar{{\cal Q}}
\rightarrow e^{-i \gamma^5 a/2 f\subs{\tiny PQ}}\,\bar{{\cal Q}}\ . 
\label{LokTraf}}
With $\gamma^\mu e^{i \gamma^5 \alpha} = e^{-i \gamma^5 \alpha}\, \gamma^\mu$ 
it is 
straightforward to show that 
under (\ref{LokTraf}) the KSVZ
Lagrangian (\ref{KSVZLag}) transforms as
\fn{{\cal L}\subs{KSVZ} 
\rightarrow
 \frac{1}{2 f\subs{PQ}} \bar{{\cal Q}}\gamma^\mu \gamma^5
{\cal Q}\, \partial_\mu a + \frac{1}{2}(\partial_\mu a)^2 - m_{\cal Q}
\bar{{\cal Q}}{\cal Q}\ . }
Obviously, the transformation (\ref{LokTraf}) generates a derivative axion 
interaction
from the kinetic ${\cal Q}$ term
in ${\cal L}\subs{KSVZ} $ (\ref{KSVZLag}), namely
\fn{{\cal L}\subs{int} = \frac{1}{2 f\subs{PQ}} \bar{{\cal Q}}\gamma^\mu 
\gamma^5
{\cal Q}\, \partial_\mu a \ .\label{IntTwo}}
In contrast with (\ref{IntOne}), this interaction contains no 
higher-order terms. The existence of two different couplings 
\fn{{\cal L}\subs{int}= -i\frac{m_{\cal Q}}{f\subs{PQ}} a \bar{{\cal Q}}
\gamma^5 {\cal Q} 
\quad \mbox{and} \quad
{\cal L}\subs{int} = \frac{1}{2 f\subs{PQ}} \bar{{\cal Q}}\gamma^\mu \gamma^5
{\cal Q}\, \partial_\mu a \label{diffcoupax}}
 raises the question of which is to use in concrete calculations.  It would 
be naive to suppose 
that both interaction Lagrangians always yield the same result because the 
pseudoscalar one is only the
leading term of an expansion in powers of $a/f\subs{PQ}$.
Therefore, it is a safe strategy to use the derivative coupling in all 
calculations.

To conclude this section we finally consider 
the effects associated with the chiral anomaly, i.e.\ the axion mass. 
We go back to
the Lagrangian (\ref{KSVZLag}) which is invariant under global $U(1)\subs{PQ}$ 
transformations (\ref{QPQtrafo}) and 
$a \rightarrow a + f\subs{PQ} \alpha$.
The associated current due to Noether's theorem  is simply
\fn{j\subs{PQ}^\mu = f\subs{PQ} \partial^\mu a - \frac{1}{2}\bar{{\cal Q}} 
\gamma^\mu \gamma^5 {\cal Q}}
with
\fn{\partial_\mu j\subs{PQ}^\mu = 0\ .}
However, the isosinglet axial current
$\frac{1}{2}\bar{{\cal Q}} \gamma^\mu \gamma^5 {\cal Q}$ has an 
Adler-Bell-Jackiw anomaly and is
therefore not conserved at the quantum level.
For the divergence of $j\subs{PQ}^\mu$ one finds explicitly 
\cite{PeskinSchroeder}
\fn{\partial_\mu j\subs{PQ}^\mu = -\frac{g^2}{32 \pi^2}\,G_a^{\mu \nu} \,
\tilde{G}^a_{\mu \nu}\ .\label{diveins} }

Recall that at the scale $f\subs{PQ}$ the PQ symmetry breaks
spontaneously and the axion arises as the associated Nambu-Goldstone boson.
As a result, the axion is a massless particle. However, at energies 
below $\Lambda\subs{QCD}$,
the axion develops a mass due to QCD effects: the heavy quark ${\cal Q}$ 
has a color anomaly. So, the axion 
interacts with gluons (Fig.\ \ref{axiongluon}), and therefore with 
quark-antiquark states.
Furthermore, the axion mass should vanish
in the limit of vanishing $u$- or $d$-quark masses.\footnote{Effects due 
to the $s$-quark are of order
$w \equiv m_u / m_s \approx 0.029$ and are neglected for the sake 
of simplicity.} With this information, Bardeen 
and Tye \cite{BardeenTye} 
constructed an anomaly free current $\tilde{j}\subs{PQ}^\mu$ 
out of $j\subs{PQ}^\mu$, 
\fn{\tilde{j}\subs{PQ}^\mu = f\subs{PQ} \partial^\mu a - \frac{1}{2}
\bar{{\cal Q}} \gamma^\mu \gamma^5 {\cal Q} 
+ \frac{1}{2}\left(\frac{m_d}{m_u + m_d}
\bar{u} \gamma^\mu \gamma^5 u + \frac{m_u}{m_u + m_d} \bar{d} 
\gamma^\mu \gamma^5 d\right) ,\label{BTC}}
where $\bar{q} \gamma^\mu \gamma^5 q,\ q = u, d$ 
are the chiral quark currents.
In the case that at least one quark mass vanishes,
the last term of expression (\ref{BTC}) is
conserved up to the chiral anomaly,\ i.e.
\fn{ \partial_\mu \left[
\frac{1}{2}\left(\frac{m_d}{m_u + m_d}
\bar{u} \gamma^\mu \gamma^5 u + \frac{m_u}{m_u + m_d}\bar{d} 
\gamma^\mu \gamma^5 d\right)
\right] \rightarrow \frac{g^2}{32 \pi^2}\,G_a^{\mu \nu}\,\tilde{G}^a_{\mu \nu}
\label{divzwei}}
for $m_u\rightarrow0$ or $m_d\rightarrow0$.
Therefore, with (\ref{diveins}), (\ref{BTC}), and (\ref{divzwei}), it 
is easy to see that the Bardeen-Tye current
is anomaly free and thus conserved in the limit of vanishing $u$- and 
$d$-quark masses, 
\fn{\partial_\mu \tilde{j}\subs{PQ}^\mu \rightarrow   0 \qquad \mbox{if}
\qquad m_u\ \mbox{or}\ m_d \rightarrow 0 .}
That means, the conservation law of the Bardeen-Tye current 
$\tilde{j}\subs{PQ}^\mu$ is explicitly 
broken by nonvanishing $u$- and $d$-quark 
masses. This fact is referred to as the partial conservation of the axial 
current (PCAC).
Using PCAC and the standard-current algebra approach \cite{currentalgebra}, 
in which the axion mass is related 
to the ``soft'' divergence 
of $\tilde{j}\subs{PQ}^\mu$, the axion mass can finally be calculated as
\fn{m_a = \frac{f_\pi m_\pi}{f\subs{PQ}} \frac{\sqrt{z}}{1 + z}\ ,}
where $z = m_u/m_d$,
$f_\pi \approx 93 \mbox{ MeV}$ the pion decay constant and 
$m_\pi = 135\mbox{ MeV}$ the pion mass. 
\section{General Axion Couplings}
For the most part, axion physics is determined by the
scale $f_a$ of PQ symmetry breaking.
According to (\ref{finallag}), the axion's couplings to photons and fermions 
are all proportional 
to the inverse PQ symmetry breaking scale or, equivalently, to the axion mass
$$\mbox{Axion-Couplings} \sim \frac{1}{f_a} \propto m_a\ .$$
In detail, however, these couplings are model dependent, i.e.\ they depend on 
the implementation of the PQ
mechanism.
\subsection{Photons\label{AxionPhotonGeneral}}
\begin{figure}[bbb]
\unitlength1mm
$\hspace{15mm}g_{a\gamma\gamma} \propto \left(\rule[-1cm]{0cm}{2cm} \right.$
\begin{minipage}{4.3cm}
\begin{picture}(40,25)
\put(-7,-10){\psfig{file=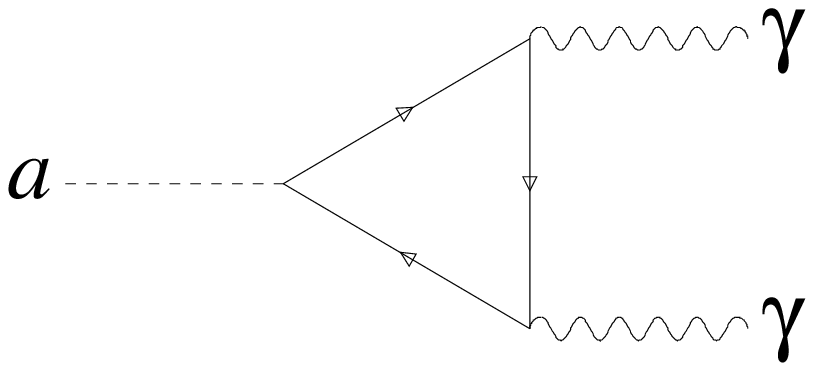,width=5.0cm}}
\end{picture}\par
\end{minipage}
$ + $
\begin{minipage}{5cm}
\begin{picture}(40,25)
\put(0,-10){\psfig{file=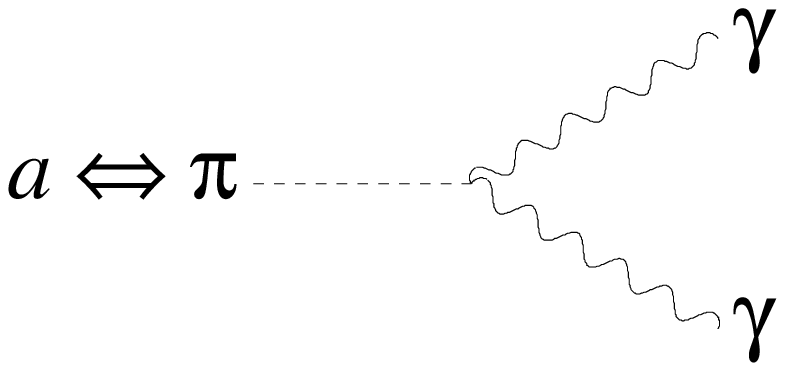,width=5.0cm}}
\put(26,15){\circle*{2.5}}
\end{picture}\par
\end{minipage}
$\left.\rule[-1cm]{0cm}{2cm} \right)$
\caption{Coupling of the axion to two photons. The 
left triangle loop is made up of
fermions carrying PQ and electric charge}
\label{neufigi}
\end{figure}

First we consider the axion's coupling to photons. An apparent reason for 
this interaction is the 
axion-pion mixing:
As a result of the electromagnetic anomaly, pions couple
to two photons, causing
an effective axion-photon coupling.
Over and above that, there is a second contribution to the axion-photon 
interaction. In general, the 
axion has Yukawa couplings to fermions which carry PQ charges.  If these PQ 
fermions also carry  electric 
charges, they couple the axion to two photons by means of a triangle loop
(Fig.\ \ref{neufigi}).
As a consequence of the axion-photon coupling,
axions decay into two 
photons with a lifetime
\begin{eqnarray}
\tau_{a\gamma\gamma}&=&\left[\frac{\alpha^2 m_a^3}{256 \pi^3 f_a^2}\,
C_{a\gamma\gamma}^{2}\right]^{-1}\nonumber\\
&=&3.53\times 10^{24}\mbox{ sec}\ m\subs{eV}^{-5}\,C_{a\gamma\gamma}^{-2}\ ,
\label{agammagamma}
\end{eqnarray} 
where $\alpha = 1/137$ is the fine-structure constant.

Let us now study some characteristic properties of the axion-photon
coupling.
If we demand CP-invariance, the effective Lagrangian for the axion-photon 
interaction can be written as
\fn{{\cal L}_{a\gamma\gamma} = -\frac{1}{4} g_{a\gamma\gamma} F_{\mu\nu} 
\tilde{F}^{\mu\nu} a = 
g_{a\gamma\gamma}\, a\, {\bf E \cdot B}\ ,}
where $F$ is the electromagnetic field strength tensor, $\tilde{F}$ its dual, 
$a$ the pseudoscalar axion field, 
and $g_{a\gamma\gamma}$ the axion-photon coupling strength with the dimension 
$(\mbox{energy})^{-1}$. The latter is given by
\fn{g_{a\gamma\gamma} = \frac{\alpha}{2 \pi f_a} \left[\frac{E}{N} - 
\frac{2(4 + z + w)}{3(1 + z + w)}\right] ,
\label{apcanalyt}}
where $\alpha = 1/137$ is the fine structure constant, $E$ the 
electromagnetic anomaly, and $N$ the color anomaly  
of 
the PQ symmetry.
They are given by
\fn{N \equiv \sum_j X_j , \qquad E\equiv 2\,\sum_j X_j\,Q_j^2\, D_j\ ,}
where $X_j$ and $Q_j$ are the PQ charges and the electric charges of the PQ 
fermions, respectively, and 
$D_j = 1$ for color singlets 
(charged leptons) and 3 for color triplets (quarks). 
Finally,  $z$ and $w$ are the mass ratios of the $u$- to the $d$-quark 
and the $u$- to the $s$-quark, respectively. 
The first term of expression (\ref{apcanalyt}) corresponds to the 
electromagnetic anomaly of the PQ fermions, 
whereas the second term is associated with the axion-pion mixing.

From (\ref{apcanalyt}) it is clear that the light quark mass ratios $z$ 
and $w$ are of great 
interest, since they determine the axion-photon coupling strength 
$g_{a\gamma\gamma}$. 
However, even though there has been
considerable effort to calculate the numerical values for $z$ and $w$, the 
results are still controversial 
and remain under active discussion. Leutwyler \cite{LeutwylerRatios} used 
chiral perturbation theory 
results for the kaon and
pion masses and extracted from those the relative sizes of $m_u$, $m_d$, 
and  $m_s$. 
He obtained the very stringent constraints 
\begin{eqnarray}
z & \equiv& m_u/m_d\ \approxeq\ 0.553 \pm 0.043 \nonumber \\
w &\equiv& m_u/m_s\ \approxeq\ 0.029 \pm 0.003\ .\label{resultLeutwyler}
\end{eqnarray}
However, alternative approaches have been made, e.g.\ using sum rules and 
numerical 
simulations 
of QCD on a lattice. These calculations lead to different results which are 
not mutually consistent.  
Therefore, it remains to be seen whether the results 
(\ref{resultLeutwyler}) are reliable. The Particle Data Group
gives the conservative range $z =$ 0.2--0.7 \cite{PDGRatio}.
Therefore, we rather use
\fn{z \approx 0.55 \pm 0.1\label{alternativz}}
than the $z$-values of expression (\ref{resultLeutwyler}).
With (\ref{axionmassformula}) and (\ref{alternativz}),
we finally obtain the axion-photon coupling in terms of the axion mass,
\begin{eqnarray}
g_{a\gamma\gamma} &=& \frac{\alpha}{2 \pi f_a} \left(\frac{E}{N}-1.93
\pm 0.08\right)
\nonumber\\
&=&\frac{m\subs{eV}}{0.52\times 10^{10} \mbox{ GeV}}\left(\frac{E}{N}-1.93
\pm 0.08 
\right) ,\label{gagamma}
\end{eqnarray}
where have used $m\subs{eV} \equiv m_a/\mbox{eV}$ and (\ref{resultLeutwyler}). 
Hereafter we use the abbreviation
\fn{C_{a\gamma\gamma} \equiv \frac{E}{N} - 1.93\ .}
In the DSVZ model and grand unified theory (GUT) models one has $E/N = 8/3$. 
This implies 
$(E/N - 1.93) \approx 0.75$.

However, in the KSVZ model
one finds $E/N = 6\, Q\subs{em}$, where 
$Q\subs{em}$ is the electric charge of the heavy quark ${\cal Q}$. 
Moreover, in more general KSVZ-type models the 
number $N$ of heavy quarks 
and their transformation
properties under $SU(3)_C$ are arbitrary. Therefore, as 
Kaplan \cite{KaplanHA} first pointed out, it is possible to construct
KSVZ-type models with $E/N = 2$, leading to a very small $C_{a\gamma\gamma}$.
In the light of the considerable
uncertainties of the light-quark ratios $w$ and $z$, it is 
even possible that 
$C_{a\gamma\gamma} = 0$. We will see in Chapter \ref{sectBounds}
that KSVZ-type axions with an accidentally suppressed photon coupling
and masses between 10 and \mbox{20 eV} are astrophysically
allowed, particles which we refer to as 
``hadronic~axions.''
\subsection{Electrons}
\subsubsection{Tree-Level Coupling to Electrons:}
In many axion models ordinary electrons carry PQ charges and have fundamental 
Yukawa couplings to the Higgs 
field. Then the tree-level coupling between 
axions and electrons has the structure of
expression (\ref{diffcoupax}). The associated coupling constant is
\fn{g_{ae}\sups{tree} = \frac{C_e m_e}{f_a} = C_e\,\, 0.85\times10^{-10}\, 
m\subs{eV} ,} 
where $C_e = X'_e / N$ is a numerical coefficient of order unity.
Here, $X'_e$ is the shifted PQ charge of the electron \cite{shiftedcharge}, 
and $N$ the color anomaly. Of course, in
KSVZ-type models one has $C_e = 0$.
\subsubsection{Radiatively Induced Coupling to Electrons:}
\begin{figure}[tb]
\unitlength1mm
\begin{picture}(70,30)
\put(30,-15){\psfig{file=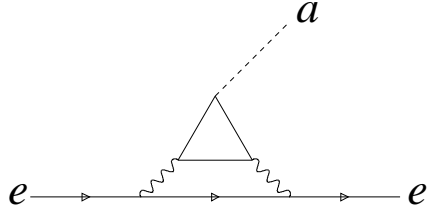,width=8.0cm}}
\end{picture}\par
\caption{Radiatively induced axion-electron coupling. \label{axelloop}}
\end{figure}

In addition to the tree-level coupling,
there exists a loop correction due 
to the axion's interaction with 
photons (Fig.\ \ref{axelloop}). Such a 
radiatively induced coupling exists
even if $X'_e = 0$.
This loop correction can be described by the effective coupling constant
\cite{shiftedcharge}
\fn{g_{ae}\sups{loop} = \frac{3 \alpha^2}{4 \pi^2}\frac{m_e}{f_a}
\left[\frac{E}{N} \ln\left(\frac{f_a}{m_e}\right)
 -  \frac{2(4 + z + w)}{3(1 + z + w)} \ln\left(\frac{\Lambda\subs{QCD}}{m_e}
\right)\right] \label{aeloop}}
with the cut-off scales $f_a$ and $\Lambda\subs{QCD} \approx 150$--$400
\mbox{ MeV}$, respectively.

Considering both the tree-level and the radiatively induced coupling, 
we obtain 
for the total 
axion-electron interaction Lagrangian
\begin{eqnarray}
{\cal L}_{ae} &=& - i \left(g_{ae}\sups{tree} + g_{ae}\sups{loop}\right) 
\bar{\psi}_e \gamma_5 \psi_e a 
\label{aegeneral}\\
 &=& - i \frac{m_e}{f_a}
\left\{
C_e + \frac{3 \alpha^2}{4 \pi^2}\left[\frac{E}{N} \ln\left(\frac{f_a}{m_e}
\right)
 -  \frac{2(4 + z + w)}{3(1 + z + w)} \ln\left(\frac{\Lambda\subs{QCD}}{m_e}
\right)\right]
 \right\}\nonumber \\
& & \times\ \bar{\psi}_e \gamma_5 \psi_e a\ .\nonumber
\end{eqnarray}
Although the loop correction is of ${\cal O}(\alpha^2)$ smaller than the 
tree-level result, it is particularly important 
for KSVZ-type axions because they have $g_{ae}\sups{tree}=0$. In this case, one
is left with the radiative coupling (\ref{aeloop}) or, numerically,
\fn{g_{ae}\sups{loop}\approx 3.4 \times 10^{-16}\ m
\subs{eV}\left[\frac{E}{N}\left(23.2 - 
\ln m\subs{eV}\right) - 12\right] .}
For hadronic axions with $E/N \approx 2$ this is $g_{ae}\sups{loop}
\approx 10^{-14}\,m\subs{eV}$.
\subsection{Nucleons}
Naturally, if ordinary quarks carry PQ charges they have fundamental 
Yukawa couplings to axions. But in astrophysical 
considerations one is confronted with energies
below the QCD scale $\Lambda\subs{QCD} \approx 150$--$400\mbox{ MeV}$. Thus, 
free quarks do not exist and one is
interested in the effective coupling of axions to nucleons. This interaction  
arises from two roughly equal 
contributions. First, there is the tree-level coupling of axions to up and 
down quarks.
Moreover,
the familiar axion-pion mixing provides a second term for the axion-nucleon 
interaction.  
The resulting effective axion-nucleon interaction can be described by a 
Lagrangian similar to that of (\ref{diffcoupax}). The axion-nucleon
coupling constant is defined as \fn{g_{aN} = \frac{C_N m_N}{f_a} ,}
where $C_N$ is the effective 
PQ charge of the nucleon. 
This charge $C_N$ is a compound parameter combining both parts of 
the axion-nucleon interaction, i.e.\
the tree-level part and the part due to the axion's mixing with pions.
For neutrons and protons it is
\begin{eqnarray}
C_p &=& (C_u - \eta) \Delta u + (C_d -\eta z) \Delta d + (C_s -\eta w) \Delta 
s\ ,\nonumber \\
C_n &=& (C_u - \eta) \Delta d + (C_d -\eta z) \Delta u + (C_d -\eta w) \Delta 
s\ ,\label{CnCp}
\end{eqnarray}
where $\eta \equiv (1+z+w)^{-1} =0.632$ with $z \equiv m_u / m_d$ and $w 
\equiv m_u / m_s$.
Furthermore,
$\Delta q$
describes the fraction of the nucleon's spin carried by the quark flavour $q$.
Numerically, one finds \cite{KeilEllis}
\begin{eqnarray}
\Delta u &=& +0.80 \pm 0.04 \pm 0.04\, , \nonumber \\
\Delta d &=& -0.46 \pm 0.04 \pm 0.04\, ,\nonumber\\
\Delta s &=& -0.12 \pm 0.04 \pm 0.04\, ,
\end{eqnarray}
where the first error is statistical, and the second is of
systematic nature.

Let us now consider KSVZ-type axions in more detail.
In spite of their vanishing couplings to ordinary quarks, their
interaction with nucleons is not zero. Actually, the axion-nucleon
coupling of KSVZ-type axions is
comparable to that which appears in models with existing
tree-level coupling to quarks, e.g.\ the DFSZ model.
The fundamental reason is the generic axion-pion mixing which exists even 
though the tree-level coupling
to quarks vanishes.
With vanishing PQ charges $C_u$, $C_d$, and $C_s$
of the $u$-, $d$-, and $s$-quarks, respectively, and with 
(\ref{CnCp}) one obtains explicitly
\fn{C_p = -0.34\quad\mbox{and}\quad C_n = 0.01 .\label{CpCn}}
These values imply
\begin{eqnarray}
g_{ap} &=& \frac{C_p m_p}{f_a} = -5.32\times10^{-8}\, m\subs{eV}\ ,\nonumber\\
g_{an} &=& \frac{C_n m_n}{f_a} = 1.57\times10^{-9}\, m\subs{eV} 
\label{gapganNum}
\end{eqnarray}
for the KSVZ Yukawa couplings.

\setcounter{equation}{0} \chapter{Hadronic Axion Bounds\label{sectBounds}}
Present bounds on the hadronic axions are reviewed; for more general
models see \cite{RaffeltPPfS}.
The possibility of novel hadronic axion bounds, based on neutron star cooling 
and accretion discs around black holes, is deferred 
to Chapter~\ref{novelimpacts}.
Limits on $f_a$ (or $m_a$) are usually based
on limits on the coupling strength to photons, electrons, and nucleons. 
These limits, in turn, are derived from 
astrophysical objects like the sun or white dwarfs where axion production would
increase stellar energy losses and thus modify the observed properties of 
stars (``energy-loss argument'').
For hadronic axions the relevant astrophysical objects are mainly 
globular cluster stars and SN 1987A.
\section{Globular Clusters}
All stars in a globular cluster have nearly the same age and equal chemical 
properties; they 
differ primarily in one single parameter, their initial mass. Particularly 
interesting for axions
are two different kinds of stars, namely those occupying the 
red-giant branch (RGB) and the horizontal 
branch (HB).
The main characteristic of the former is their degenerate helium core, 
whereas the
latter are 
characterized by a helium-burning core. In both kinds of stars the core is 
surrounded by a 
hydrogen burning shell.
In the interior of HB and RGB stars axions could be produced by different 
mechanisms, leading 
to 
bounds on the axion-photon and the axion-electron coupling.
\subsection{Photon Coupling}
In globular-cluster stars photons can be transformed into axions by 
means of the axion-two photon interaction. 
This effect is referred to as Primakoff conversion.
It takes place in both the HB and RGB star cores, but, as a consequence of 
the different
core densities, Primakoff conversion is much more effective in HB stars.
Therefore, an additional axionic 
energy loss would shorten the lifetime of HB stars, leading to a 
reduced HB/RGB number ratio, 
in contradiction to observations.
These considerations lead to the 
bound \cite{RaffeltSaLfFP}
\fn{g_{a\gamma\gamma} \simlt\, 0.6\, \times 10^{-10}\mbox{ GeV}^{-1}\ .
\label{gagaaamma}}
With (\ref{gagamma}) one obtains
\fn{C_{a\gamma\gamma}\,m\subs{eV} \simlt 0.3\ ,\label{boundongae}}
where again
$m\subs{eV}= m_a / \mbox{eV}$ and $C_{a\gamma\gamma} = (E/N -1.93)$.
Obviously, 
the limits on $C_{a\gamma\gamma}$ and $m_a$ are not independent.
In particular, for hadronic axions we have
$C_{a\gamma\gamma} \ll 1$, and no meaningful
limit on $m_a$ may be derived.
\subsection{Electron Coupling}
Another constraint may be obtained by 
considering the axionic energy loss via the 
bremsstrahlung process $e^- +(A,Z) \rightarrow e^- +(A,Z) + a$.
The consequence of this energyloss would be
a delay of the helium flash in the red-giant core, resulting in
a greater core mass at helium ignition. 
A comparison 
between the
predicted and observed core masses at the helium flash yields 
\cite{RaffeltSaLfFP}
\fn{g_{ae} \simlt\, 2.5 \times 10^{-13}\ .}
In the case of hadronic axions, $g_{ae}$ is given by the radiatively induced 
coupling, implying the condition
\fn{m\subs{eV}(C_{a\gamma\gamma} + 1.3) \simlt 35 .\label{bbbgae}}
As a result of the SN 1987A
bound, which will be discussed below, the mass of hadronic axions
is restricted to the range between 10 and 20\mbox{ eV}. 
This requires $C_{a\gamma\gamma}$ to be less than 0.45--2.2,
which can be easily provided in
KSVZ-type models. 
\section{SN 1987A}
Axions could be produced via nucleon-nucleon axion bremsstrahlung
$N + N \rightarrow N + N + a$ in the core of a supernova. As a
consequence, the most important 
bound on hadronic axion 
masses can be derived from the SN 1987A by considering the strength
and duration of the observed neutrino signal at the Kamiokande~II 
(KII) and the 
Irvine-Michigan-Brookhaven (IMB)
water-Cherenkov detectors.
Depending on their masses $m_a$, axions either escape
freely or are radiated from an axion sphere. This energy loss can 
have three different
observable effects: first, it can significantly shorten the observed 
neutrino burst, 
second, it can cause
additional counts  in the neutrino detectors, and finally, emitted axions 
could alter the extragalactic background light via their 
radiative decay.
\subsection{Nucleon Coupling}
Axions produced in the SN core 
compete with the standard neutrino cooling 
channel, i.e.\
they remove energy from the neutrino signal, implying 
a shortening of the observed 
neutrino signal.
The axion luminosity $L_a$ depends on the axion-nucleon 
coupling $g_{aN}$  and 
therefore indirectly on the
axion mass $m_a$. Assuming that the mean free path of axions is larger than 
the size of the core, axions
are able to escape freely. 
In this instance the axion luminosity $L_a$ 
increases with increasing axion mass:
the greater $m_a$, the stronger $g_{aN}$, and thus the axion emission rate. 
On the other hand, if axions interact too strongly, they are trapped inside 
the 
core and are emitted
from an axion sphere, similar to the well-known 
concept of a neutrino sphere. A simple analytical 
model
\cite{AxionTrappModa,AxionTrappModb} showed that the axion luminosity in the 
trapped 
regime varies 
as $L_a \propto m_a^{-16/11}$, i.e.\ $L_a$ decreases with increasing axion 
mass $m_a$. 
Naturally, beyond some large coupling  $g_{aN}$, axions 
will be trapped so effectively 
that their impact on
the SN cooling is again negligible. Summarizing these arguments, we 
can conclude 
that axions must
either couple sufficiently weakly or sufficiently strongly to nucleons 
to avoid 
a conflict with
the observed neutrino signals at KII and IMB. Accurate calculations 
\cite{KeilRaffeltSN,KeilEllis,RaffeltSaLfFP} yield the excluded range
\fn{3 \times 10^{-10}\, \simlt\, g_{aN}\, \simlt\, 3 \times 10^{-7}\ .}
Assuming a proton fraction of 0.3 inside the SN core, one can calculate an 
effective
axion-nucleon coupling $C_{aN} \equiv \sqrt{0.3 C_{ap}^2 + 0.7 C_{an}^2} 
\approx 0.2$, 
where we have used
(\ref{CpCn}). With $g_{aN}=C_N m_N / f_a$ one finally obtains
\begin{eqnarray}
0.01\mbox{ eV}\, \simlt& m_a &\simlt\, 10\mbox{ eV}\ ,\nonumber \\
6 \times 10^{8}\mbox{ MeV}\, \simlt & f_a & \simlt\, 6 \times 10^{11}
\mbox{ MeV} 
\label{SNexclusion}
\end{eqnarray}
for the SN 1987A exclusion parameters of hadronic axions.

But even axions which are allowed by the energy-loss argument
can have measurable 
effects, 
as they might
produce additional counts in the water-Cherenkov detectors. 
Engel et al.\ \cite{EngelSeckelHayes}
showed that such heavy hadronic axions are able to induce nuclear excitations 
in oxygen,
$a + {^{16}\mbox{O}} \rightarrow {^{16}\mbox{O}}^{\ast}$.
After that, nuclear deexcitations produce 
$\gamma$-rays which can then be detected. Hence, one can estimate a range
between
\begin{eqnarray}
20\mbox{ eV}\, \simlt& m_a &\simlt\, 20\mbox{ keV}\ ,\nonumber \\
3 \times 10^{5}\mbox{ MeV}\, \simlt & f_a & \simlt\, 3 \times 10^{8}
\mbox{ MeV} 
\label{SNCounter}
\end{eqnarray}
for excluded hadronic axion masses.
\subsection{Photon Coupling}
If axions had been emitted from SN 1987A, they might have 
produced a flux of
$\gamma$-rays due to their radiative decay mode $a\rightarrow 2 \gamma$.  
Hadronic axions with $m_a \sim 10\mbox{ eV}$ should have been emitted with a 
fluence of approximately
$f_a \approx 5.7 \times 10^{10}\, m\subs{eV}^{-12/11}\mbox{ cm}^{-2}$ 
\cite{AxionTrappModa}.
On condition that the axion-photon coupling $g_{a\gamma\gamma}$
is strong enough, the resulting
$\gamma$-fluence $f_\gamma$ 
could have been detected by the gamma-ray spectrometer (GRS) on 
the Solar Maximum Mission (SMM) satellite. The fact that the GRS did not 
detect any 
signal above the instrument
background sets an upper limit to the expected $\gamma$-fluence $f_\gamma$ 
\cite{RaffeltSaLfFP, KolbTurnerRadLim},
\fn{f_\gamma \approx 10^{-9}\,C_{a\gamma\gamma}^2 m_a^{58/11} \simlt 0.4\ .}
The SN 1987A  bound derived in the previous section precludes axion masses 
smaller than
$10\mbox{ eV}$ and greater than $20\mbox{ eV}$.
Therefore, if we use \mbox{20 eV} as a maximum 
value for the mass we obtain the conservative limit
\fn{C_{a\gamma\gamma} \simlt 7 ,}
which is less stringent than (\ref{boundongae}).
\section{Gamma-Ray Background Limits}
Nontrivial constraints come from the axionic contribution to the 
extragalactic 
background light. 
As we will see in Chapter \ref{relicaxions}, hadronic axions could 
have been produced 
in the early universe.
These relic axions could have left their mark due their radiative decay 
mode 
$a\rightarrow 2 \gamma$,
provided that the axion-photon coupling $g_{a\gamma\gamma}$ is sufficiently 
strong.

Overduin and Wesson \cite{OverduinWesson} considered the diffuse 
extragalactic background 
light and searched
for axion-decay photons. The absence of a signal implies
\fn{C_{a\gamma\gamma} < 0.323,\, 0.05,\,
0.015\ \ \  \mbox{for axion masses}\ \ \ m\subs{eV} = 5.3,\, 8.6,\,13\ .
\label{relybound}}
These limits give strong constraints on the 
axion-photon coupling, but they
can be fulfilled assuming an accidental cancellation 
$C_{a\gamma\gamma} = 0$.

Furthermore, Ressel \cite{TedRessell} searched for photon emission lines 
in the center of three 
clusters of galaxies,
i.e.\ he considered the flux from a particular region of the sky rather 
than the whole sky. 
He obtained
the bounds $C_{a\gamma\gamma} <$ 0.12, 0.059, 0.029,
0.024, 0.012, 
and 0.008
for $m\subs{eV}=$ 3.5, 4.0, 4.5, 5.0, 6.0, and 7.5, respectively. 
However, masses less then 
$10\mbox{ eV}$
are already ruled out as a consequence of the energy loss argument 
applied to SN 1987A.
\section{Big Bang Nucleosynthesis (BBN)}
The bounds of the previous sections have been frequently 
discussed so that we 
just gave a brief
summary. However, the possible impact of hadronic axions on
BBN has not been 
investigated in such
detail, justifying a closer look.

It will be shown in the next chapter that hadronic axions with masses 
$m_a =$ 10--20~eV 
would have come
into thermal equilibrium after the quark-hadron phase transition. 
At temperature 
$T_D \sim 60\mbox{ MeV}$ they
decouple and are thus highly relativistic. 
The fact that hadronic axions might have an energy density $\rho_a$ comparable 
to that of a light
neutrino species would affect the outcome of BBN
which takes place at
about 0.05~MeV. During that epoch the universe is radiation dominated, and
the expansion rate $H$ depends on
the total energy density $\rho$ according to the Friedmann equation,
\fn{H^2 \equiv \left(\frac{\dot{R}}{R}\right)^2 = 
\frac{8 \pi}{3} G_N \rho .}
If one takes the existence of relativistic axions at $T \approx 1\mbox{ MeV}$ 
into account, 
the expansion parameter is increased due 
to the axion's energy density $\rho_a$. 
An increase of
$H$ means that the
all-important 
neutron-proton ratio $n/p$, which is regulated by the weak 
interactions,
freezes out at an earlier time, 
when $n/p$ was larger. Therefore, taking hadronic 
axions into account, more $^4\mbox{He}$ is synthesized than in the standard 
scenario.

In order to estimate the influence of thermally produced hadronic axions on 
BBN, 
one compares their energy density $\rho_a$ with that of
a light neutrino species, e.g.\ the electron neutrino. Both particles are 
relativistic so that
\fn{\Delta N \equiv \left(\frac{\rho_a}{\rho_\nu}\right) = \frac{4}{7}
\left(\frac{T_a}{T_\nu}\right)^4 ,
\label{aabbaa}}
where the factor $4/7$ is due to different phase-space
distributions and different spins of 
neutrinos and axions. This equation 
means that the axion's energy density is equivalent to an effective number
$\Delta N$ of additional light neutrinos.
When axions decouple at $T_D \sim 60\mbox{ MeV}$, $T_a$ equals $T_\nu$ 
and thus $\Delta N = 4/7$.  
Below $T_D$ the axion temperature scales as 
$T_a \propto R^{-1}$. The light neutrino decouples later at 
$T \approx 1\mbox{ MeV}$
and $g_\ast\,T_\nu^3\,R^3$ remains constant. In the standard scenario,
the total number of effective, relativistic spin degrees of freedom
does not change until
BBN takes place, so that $T_nu \propto R^{-1}$. Therefore, we obtain
for $\Delta N$ 
at the time of BBN
\fn{\Delta N\subs{BBN} \equiv \left.\left(\frac{\rho_a}{\rho_\nu}\right)
\right|_{T=T\subs{NS}}
= \frac{4}{7} .\label{neulimit}}
This result has now to be compared with the standard model of BBN, something
that turns out to 
be anything but simple:
In accordance with the standard BBN scenario, the mass fraction of helium,
conventionally referred 
to as
$\mbox{Y}\subs{p}(^4\mbox{He})$, depends not only on $N\subs{BBN}$ but also 
on the 
uncertainties of the 
baryon-to-photon ratio $\eta \equiv n_B / n_\gamma$. 
The predicted
helium abundance must then be compared with the observed $^4\mbox{He}$ 
abundance,
a quantity that is difficult to estimate, implying 
that an upper constraint on 
$\Delta N\subs{BBN}$ is afflicted with significant uncertainties.
Different authors obtain different results: Olive et al.\ 
\cite{walkeretal} found the restrictive limit of 
$\Delta N\subs{BBN} \le 0.3$. However, Kernan and Sarkar 
\cite{kernansarkar} provide the conservative bound
$\Delta N\subs{BBN} \le 1.53$.
If we use that value, no danger is ahead as far as 
hadronic axions are concerned because our result (\ref{neulimit})
is within the above constraint.
A review of BBN, $\Delta N\subs{BBN}$-limits and related issues 
can be found in \cite{ssakar}.
In a nutshell, due to the lack of reliable data, it is premature to 
infer from BBN-based arguments 
that 
a hadronic axion with $f_a \sim 10^6\mbox{ GeV}$ can not exist.
\section{The Hadronic Axion Window}
In summary, there 
is a gap of allowed axion masses between 10 and 
$20\mbox{ eV}$. 
This is a 
result of the fact that
the two different SN 1987A exclusion regions---on the one hand 
the ``too much energy loss'' region (\ref{SNexclusion}), 
on the other the ``too many events in detectors'' 
region (\ref{SNCounter})---do not overlap. 
This 10--20\mbox{ eV} window is open because 
the globular-cluster bounds (\ref{gagaaamma}), (\ref{bbbgae}), and the
$\gamma$-ray background observation limits (\ref{relybound}) are 
of no importance if $C_{a\gamma\gamma}$ is accidentally suppressed.
According to our discussion in Sect.\ 
\ref{AxionPhotonGeneral} this is well possible.
Consequently, we are 
not able to rule out
KSVZ-type axions with a strictly suppressed coupling to photons.

\setcounter{equation}{0} \chapter{Relic Axions\label{relicaxions}}
Thermal production of hadronic axions in the
early universe is discussed. We 
show that these particles are highly relativistic
when they freeze out, implying that \has behave like HDM. Moreover, we
estimate the present axion density and find that it
is comparable with that required in mixed dark matter scenarios.
\section{Thermal Production in the Early Universe}
If axions were produced in the early universe, they could constitute
dark matter. In principle, there are two different production
mechanisms. First, axions can arise as a result
of the usual ``freeze out process.''
However, this thermal production is only meaningful
if
the axion couplings are sufficiently strong, 
implying that $m_a$ must be greater than
several $10^{-2}\mbox{ eV}$ \cite{turnersolo}. 
Furthermore, there are two nonthermal production mechanisms, 
according to different cosmological scenarios:
If inflation occurred after the PQ symmetry breaking or if
$T\subs{reheat} < f_a$, axions are produced by means of the 
``misalignment mechanism.'' Before the QCD phase transition,
the parameter $\bar{\Theta}$ is not at its CP-conserving
minimum $\bar{\Theta}=0$, but somewhere between 0 and $\pi$. Later, at
a temperature around $T \sim \Lambda\subs{QCD}$, the
$\bar{\Theta}$-field
rolls toward $\bar{\Theta}=0$, resulting in coherent oscillations
which correspond
to a condensate of zero-momentum axions.
On the other hand, if the universe did not inflate at all or
if inflation occurred before the PQ symmetry breaking with 
$T\subs{reheat} > f_a$,
a network of cosmic axion strings forms at $T \sim f_a$ by the Kibble 
mechanism, gradually
decaying into (massless) axions. At lower
temperature $T \sim \Lambda\subs{QCD}$, these axions acquire a small mass
and~become~nonrelativistic.

As a consequence of the axion's Nambu-Goldstone nature,
both the misalignment and the string-decay picture 
predict that the axion energy density $\rho_a$ is
proportional to $m_a^{-1.175}$, implying a lower limit on the
axion mass around $10^{-3}\mbox{ eV}$.
Moreover, in the mass range between approximately
$10^{-5}$ and $10^{-3}\mbox{ eV}$, nonthermally produced axions
would be an ideal candidate for CDM.
However, due to unknown initial conditions on $\bar{\Theta}$, and 
uncertainties in the quantitative treatment of the string mechanism,
the axion mass density $\rho_a$
is not straightforward to calculate so
that the function $Omega(m_a)$ remains uncertain. 

In this work we are mainly concerned with hadronic axions. 
For these particles, the nonthermal production takes 
place at a temperature 
where the axions are still in contact with the thermal bath of the universe,
implying that nonthermal mechanisms do not take place. 
Therefore, we will now focus on the thermal production of hadronic axions.

The interactions of axions with the thermal bath are all 
of the general form \fn{a + X_1 \leftrightarrow X_2 + X_3 ,
\label{genprodform}}where
$X_i$, $i=$1, 2, 3 are particles of the primordial heat bath.
Then, axion production 
is described by the covariant Boltzmann equation, and we may write
for the number of axions in a comoving volume $Y=n_a/s$ \cite{EarlyUniverse}
\fn{\frac{dY}{dx} = -\frac{\Gamma\subs{abs}}{x H}\, (Y - Y\subs{EQ}),
\label{boltz}}
where $Y\subs{EQ} = n_a\sups{EQ}/s$ is the equilibrium number of
axions in a comoving volume, $H$ the expansion rate of the universe, and 
$ \Gamma\subs{abs} = n_1 \langle\sigma 
v\rangle$ is the thermal averaged axion absorption rate of the
process (\ref{genprodform}). Furthermore,
$s=S/R^3 = (2\pi^2/45)g_\ast T^3$ is the entropy density, where
$g_\ast$ is the total number of effective spin degrees of freedom of all 
relativistic bosons and fermions that are
in thermal equilibrium at the given temperature.
Finally, 
we used the scaling parameter $x\equiv m_N /T$ with the
nucleon mass $m_N$ and the ambient temperature $T$. 

\begin{figure}[tb]
\unitlength1mm
\begin{picture}(70,60)
\put(10,-2){\psfig{file=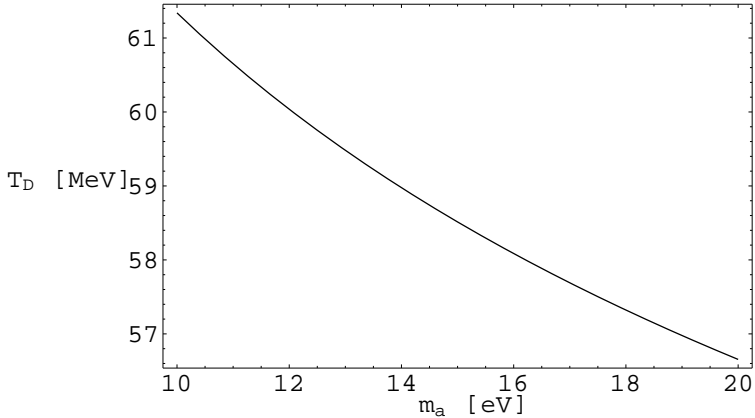,width=10.0cm}}
\end{picture}
\caption{Decoupling temperature $T_D$ of hadronic axions as 
a function of their mass $m_a$. \label{freezy}}
\end{figure}

Axions decouple from the thermal bath when $Y$ does not change anymore,
i.e\ $dY/dx \approx 0$. According to
equation (\ref{boltz}), this freeze out happens when 
the ratio $\Gamma\subs{abs} / H$ becomes small, 
$\Gamma\subs{abs}\simlt H$. 
Let us now estimate the decoupling temperature $T_D$ for hadronic 
axions, 
i.e.\ the temperature $T_D$ at which
$\Gamma\subs{abs} \approx H$. Under the assumption $T<\Lambda_{QCD}\approx 
300\mbox{ MeV}$ nucleons already exist in the 
universe and a possible realization of (\ref{genprodform}) is the
axion-pion conversion \fn{a + N \leftrightarrow \pi + N .}
With the couplings
$g_{aN} = m_N/f_a,\, g_{\pi N}\approx m_\pi^{-1}$, and a factor $m_N^{-1}$ 
for the nonrelativistic 
nucleon propagator one can estimate the interaction cross section for 
this process
\fn{\langle\sigma v\rangle \approx \left(\frac{m_N}{f_a}\frac{1}{m_\pi}
\frac{1}{m_N} T\right)^2 = 
\frac{T^2}{f_a^2 \,m_\pi^2}}
and therefore
\fn{\Gamma\subs{abs} =n_N \langle\sigma v\rangle \approx \frac{n _N T^2}
{f_a^2 \,m_\pi^2}\ .\label{gammaabsform}} 
In the case of nonrelativistic nucleons with a chemical potential 
$\mu \ll T $ the number density is given by
$n_N \approx (m_N T)^{3/2}\, e^{-x}$. Furthermore, in the radiation 
dominated early 
universe one has
\fn{H = 1.67\, g_\ast^{1/2} \frac{m_N^2}{m\subs{Pl}\, x^2} ,
\label{exprateform}} where $m\subs{Pl}$ is the Planck mass.
For the relevant temperatures, the particles
contributing to  $g_\ast$ are
$\gamma$, $e^+$, $e^-$, 
$\nu_e$,
$\nu_\mu$, $\nu_\tau$, $\bar{\nu}_e$, 
$\bar{\nu}_\mu$, $\bar{\nu}_\tau$, and $a$,
implying
$g_\ast = 11.75$ so that we obtain
\fn{\frac{\Gamma\subs{abs}}{H}\approx 0.60 \,\frac{m_N^3\, m\subs{Pl}}{f_a^2\,
m_\pi^2\,g_\ast^{1/2}}\,
x^{-3/2}e^{-x} \approx 2.7 \times 10^6\, m\subs{eV}^2\,
x^{-3/2}e^{-x}\ ,
\label{ratioGH}}
with $m\subs{eV} \equiv m_a/\mbox{eV}$. 
The decoupling temperature $T_D$ is reached when this ratio equals 1.
Fig.\ \ref{freezy} shows $T_D$ as a function
of hadronic axion masses between $10\mbox{ eV}$ and $20\mbox{ eV}$,
i.e.\ for the parameter range of the \ha window. 
We see that the decoupling temperature
is approximately
\fn{T_D \approx \mbox{57--61}\mbox{ MeV}\ .\label{entkopplung}}
Although this calculation was only a rough estimate, there is no doubt
that hadronic axions are highly relativistic when they freeze out.
Hence, their equilibrium number density is $n_a = \zeta(3) T^3 /\pi^2$ so that
we find
\fn{Y\subs{EQ}= \frac{n_a^{EQ}}{s}=\frac{45\, \zeta(3)}{2 \pi^4}\frac{1}
{g_\ast} \approx 0.024 \label{eqvalueaxion}}
for the equilibrium number per comoving volume.
To estimate the present abundance of 
relic axions, we return to the Boltzmann equation (\ref{boltz}) which
can be solved easily because 
$Y\subs{EQ}$ is constant.
With the boundary condition $Y(0)=0$, i.e.\ initially no axions were 
present, one finds
\fn{Y(x) =Y_{EQ}\left(1 - \exp\left[-\int_0^{x} \frac{\Gamma\subs{abs}}{x' H} 
dx'\right]
\right) .\label{solu}}
Strictly speaking, we
would have to integrate from the 
very beginning
($x=0$) to the present ($x = x\subs{today}\equiv x_0$) 
to find the number of axions in the universe. 
However, today 
axions are decoupled, and the present
ratio $\Gamma\subs{abs}/H$ in
(\ref{solu}) is extremely small. Thus we may integrate forward to 
$x=\infty$ rather then to $x = x_0$.
Moreover, we are interested in the time after the QCD phase transition
which occurs roughly at $x\subs{QCD}
\equiv m_N/\Lambda\subs{QCD} \approx 5$
so that
\fn{Y(x_0) = Y_{EQ}\left(1 - \exp\left[-\int_{x\subs{QCD}}^{\infty}
\frac{\Gamma\subs{abs}}{x' H} 
dx'\right]\right) .}
Together with expression (\ref{ratioGH}) and hadronic axion masses 
$m_a={\cal O}(10\mbox{ eV})$, it is easy to see that the exponential
function $\exp[-\int\ldots]$ can be neglected.
Hence we find that the
present number of axions per
comoving volume is identical with its value at freeze out, i.e.\
\fn{Y(x_0) = Y\subs{EQ}(x_D)\approx 0.024 .\label{exactvalueofaxi}}
Now we can calculate the present number density $n_a^0$ of relic hadronic 
axions. 
With $s_0 = 2970\mbox{ cm}^{-3}$
for the present entropy density, we find
\fn{n_a^0 = Y(x_0)\, s_0 \approx 71\mbox{ cm}^{-3}\ .\label{axionden}}
One can compare this result with the number 
density $n_\nu^0 = 115\mbox{ cm}^{-3}$ 
of relic electron neutrinos and obtains
\fn{\left(\frac{n_a^0}{n_\nu^0}\right)\approx 0.6 .\label{ratioaxnu}}

Our calculation of the hadronic axion decoupling temperature 
(\ref{entkopplung}) and its present
number density, (\ref{axionden}) or (\ref{ratioaxnu}), is just an estimate 
because
we have only considered the axion-pion conversion process 
$a + N \leftrightarrow \pi + N$. A more
careful approach includes an accurate computation of all relevant processes. 
First, these are the 
reactions
\fn{\pi^0 n \leftrightarrow a n,\ \pi^0 p \leftrightarrow a p,\ 
\pi^+ n \leftrightarrow 
a p,\ \pi^- p 
\leftrightarrow a n .\label{procone}}
In addition, there are reactions involving only pions rather than nucleons 
and pions,
\fn{\pi^+ \pi^- \leftrightarrow a \pi^0,\ \pi^0 \pi^\pm 
\leftrightarrow a \pi^\pm .
\label{proctwo}}
These processes were evaluated by Chang and Choi \cite{ChangChoi}. 
They assert that
the pure pion processes (\ref{proctwo}) dominate over the mixed ones 
(\ref{procone}) for temperatures up to 150\mbox{ MeV},
which is above our estimated decoupling temperature $T_D \sim 60\mbox{ MeV}$. 
Therefore, the total interaction rate including
all relevant processes is greater than our approximation, implying that
the ratio $\Gamma/H$ equals 1 at lower temperatures, i.e.\
axion decoupling
occurs at temperatures below $T_D \sim 60\mbox{ MeV}$. In fact,
Chang and Choi found
$T_D \approx 30$--$50\mbox{ MeV}$.
However, axions are relativistic particles when they freeze out, 
and thus the present number density $n_a^0$ of relic axions is quite
insensitive to
the exact value of the 
decoupling temperature.
The reason is that $n_a^0$
in (\ref{axionden}) depends
only slightly upon $T_D$, through $\left.g_{\ast}\right|_{x=x_D}$.  
However, both for $T_D \sim 60\mbox{ MeV}$ and
$T_D \approx 30$--$50\mbox{ MeV}$,
the effective number of spin degrees of freedom $g_\ast$ is
given by 11.75. Therefore,
our estimate (\ref{exactvalueofaxi}) 
agrees with the exact calculation
taking all interaction processes into account, which
yields $Y(x_0)\approxeq 0.02$ \cite{Moroietal}.
\section{Hadronic Axions as Hot Dark Matter}
The existence of thermally produced hadronic axions,
which are highly relativistic when they freeze out, means that
these particles contribute to cosmic hot dark matter (HDM)\@.
The present hadronic axion number density (\ref{axionden}) can be translated
into the axion contribution
to the present mass density, one obtains
\fn{\Omega\subs{HDM}\, h^2 = 0.007\, m\subs{eV} ,}
where $h$ is the Hubble constant $H_0$
in units of 100 km  $\mbox{sec}^{-1}$
$\mbox{Mpc}^{-1}$. Numerically, one finds with $h\approx$ 0.6--0.8 and 
$m_a \approx 10$--$20\mbox{ eV}$
\fn{\Omega\subs{HDM} \approx \mbox{0.1--0.4} .\label{massdensHDM}} 
This is the right ballpark for a HDM component required in mixed 
dark matter scenarios,
\cite{novosyadlyj,ValKahNov}. 

\setcounter{equation}{0} \chapter{Axions in a Nuclear Medium \label{nucmedres}}
It is the nature of hadronic axions that they
couple primarily to nuclear matter. Therefore, suitable
astrophysical objects for investigating their properties are those which
consist mainly of nucleons, e.g.\ supernovae, neutron stars, and
accretion discs. The impact of
axions on SN 1987A was already discussed in Chapter \ref{sectBounds}, while
consequences of hadronic axions for
neutron star and accretion disc physics will be explored in Chapter
\ref{novelimpacts}.
The extent to which these objects 
are affected by axions
depends strongly
on the axion's emission and absorption properties.
Therefore, we devote the present chapter to the
axion emission rates and mean free path
in neutron stars and accretion discs.
\section{Axion Emission in a Nuclear Medium\label{AEIANM}}
\subsection{Nucleon-Nucleon Axion Bremsstrahlung}
As a consequence of the axion-nucleon coupling,
axions can be produced in a nuclear medium via the nucleon-nucleon axion
bremsstrahlung process $N + N \rightarrow N + N + a$, where
$N$ can be a proton $p$ or a neutron $n$. For this
production mechanism to
be possible, one has to employ a spin-dependent nucleon-nucleon 
potential. An appropriate ansatz is the 
one-pion exchange (OPE) potential, which
allows a simple calculation of the bremsstrahlung process.
It should be stressed that using an OPE potential is just
an approximation to the real
nucleon interactions in hot and dense matter.
However, since our goal is a simple estimate of the axion's influence
on \ads and neutron stars, the 
OPE potential is a suitable assumption.
Furthermore, we assume nonrelativistic nucleons.
This is justified because the temperatures in \nss and
\ads are around 100~keV and 10~MeV, respectively.

\begin{figure}[tb]
\unitlength1mm
\begin{picture}(70,35)
\put(7,-30){\psfig{file=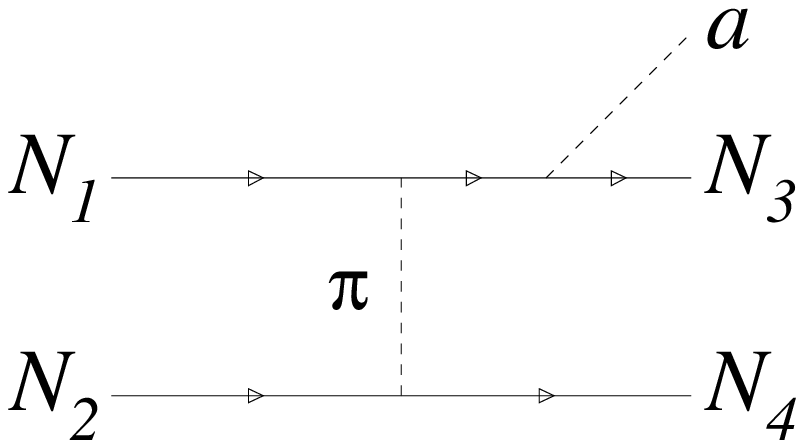,width=7.0cm}}
\put(65,-30){\psfig{file=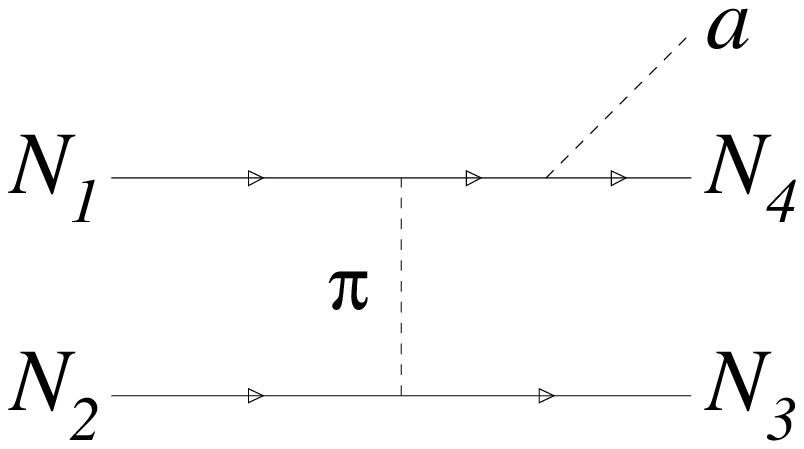,width=7.0cm}}
\end{picture}
\caption{One possible Feynman diagram and its exchange graph
for nucleon-nucleon axion bremsstrahlung. Altogether there are
eight graphs, four direct and four exchange diagrams.}
\label{bremsstrahlFig}
\end{figure}

In the OPE approximation one has in total eight Feynman diagrams for
the bremsstrahlung process: four direct diagrams with
the axion attached to each nucleon line, and four exchange graphs 
each with $N_3 \leftrightarrow N_4$
(Fig.~\ref{bremsstrahlFig}).
For the ``pure'' processes
$n n \rightarrow n n +a$ and $p p \rightarrow p p +a$ with only
neutrons and protons, respectively, 
the spin-summed squared matrix element 
is found to be \cite{brinkmann}
\begin{eqnarray}
\sum_s |{\cal M}|_{NN}^2 = 
\frac{256 \pi^2 \alpha_\pi^2}{3 m_N^2}\,g_{aN}^2
\Bigg[
\left(\frac{{\bf k}^2}{{\bf k}^2 + m_\pi^2}\right)^2  + 
\left(\frac{{\bf l}^2}{{\bf l}^2 + m_\pi^2}\right)^2  \nonumber \\
 {}  +  \frac{{\bf k}^2\, {\bf l}^2 - 3 ({\bf k \cdot l})^2}{({\bf k}^2 
+ m_\pi^2)({\bf l}^2 + m_\pi^2)}\Bigg]
 ,\label{matrixcomplete}
\end{eqnarray}
where $g_{aN} = C_N\, m_N / f_a$, $N= p,n$ is the 
axion-nucleon coupling constant and
$\alpha_\pi = (f 2 m_N/m_\pi)^2/4\pi$ the pion-nucleon ``fine-structure 
constant'' with the pion-nucleon coupling
$f \simeq 1.05$.
Furthermore, we have ${\bf k} = {\bf p}_2 - {\bf p}_4$ and 
${\bf l} = {\bf p}_2 - {\bf p}_3$ with the nucleon's momenta 
${\bf p}_i$.
For the ``mixed'' process $n p\rightarrow n p + a$ one obtains
\begin{eqnarray}
\sum_s |{\cal M}|_{np}^2 &=& \frac{256 \pi^2 \alpha_\pi^2}{3 m_N^2}\,
\frac{(g_{an}+g_{ap})^2}{4}
\left[
2\left(\frac{{\bf l}^2}{{\bf l}^2 + m_\pi^2}\right)^2 -
\frac{4\,({\bf k \cdot l})^2}{({\bf k}^2 + 
m_\pi^2)({\bf l}^2 + m_\pi^2)}
\right]
\nonumber\\
& & {}+ \frac{256 \pi^2 \alpha_\pi^2}{3 m_N^2}\,\frac{g_{an}^2+g_{ap}^2}{2}
\Bigg[
\left(\frac{{\bf k}^2}{{\bf k}^2 + m_\pi^2}\right)^2 +
2\left(\frac{{\bf l}^2}{{\bf l}^2 + m_\pi^2}\right)^2 \nonumber \\
& &\hspace{5cm}{} + 2 \frac{{\bf k}^2\, {\bf l}^2 
- ({\bf k \cdot l})^2}{({\bf k}^2 + 
m_\pi^2)({\bf l}^2 + m_\pi^2)}
\Bigg] .\label{matrixcompleteNP}
\end{eqnarray}
The larger coefficients compared to (\ref{matrixcomplete})
are a consequence of the
stronger couplings of charged pions to nucleons.
To obtain the axion energy loss rate per unit volume, 
one has to perform the phase-space integration
\begin{eqnarray}
Q_a =\int \prod_{j=1}^{4,a} \frac{d^3 p_j}{2 E_j (2 \pi)^3} E_a\,S
\sum_s |{\cal M}|^2 \,  f_1 f_2 (1-f_3)(1-f_4) \nonumber
\\ \times\ (2 \pi)^4\, \delta^{(4)}(P_1 + P_2 - P_3 - P_4 - P_a) ,
\label{PhaseSpace}
\end{eqnarray}
where $P_j$, $j = 1,\ldots,4$ are the four-momenta of the external 
nucleons, $P_a$ the one of the axion, and $f_j$, $j = 1,\ldots,4$
the phase-space distributions of the nucleons. The factor $S$ takes
identical particles in the initial and final states into account:
$S=1/4$ for the pure, and $S=1$ for the mixed processes.
The Bose stimulation factor for the final-state axion
$(1+f_a)$ was neglected because
we will see that these particles always escape freely.

To be able to perform the phase-space 
integration one needs to specify the phase-space distributions $f_i$
of the nucleons.
For accretion discs and \nss this is a difficult task.
Therefore, as a first approximation we describe the nuclear medium
in these objects
as an ideal gas of protons, neutrons, and 
electrons,
where neutrons and protons have approximately
the same mass of $m_N \approx 940\mbox{ MeV}$. 
Then, the phase-space occupation function of protons $(i=p)$ and neutrons
$(i=n)$ is given by the Fermi-Dirac distribution
\fn{f({\bf p}_i) = \frac{1}{\exp[(E_i({\bf p}_i) - \mu_i)/T] + 1}
\label{fermidiridist}\, .}
As the nucleons are nonrelativistic, it is appropriate to
use the nonrelativistic kinetic energies 
$E\sups{kin}_i = E_i({\bf p}_i) -m_i\approx {\bf p}_i^2 / 2\, m_i$ 
and the nonrelativistic chemical potentials 
$\hat{\mu}_i = \mu_i - m_i$.

The integration in (\ref{PhaseSpace})
simplifies considerably for degenerate or nondegenerate conditions, limits
which pertain to \nss and accretion discs, respectively. For the latter,
characteristic temperatures and densities are
around $T \simlt \mbox{10 MeV}$ and
$\rho_B \approx 10^{12}\mbox{ g cm}^{-3}$, respectively,
so that we find for the nucleon
(nonrelativistic) Fermi energy 
\fn{\varepsilon_{F,N}= p_{F,N}^2 / 2\,m_N = (3 \pi^2 Y_N 
n_B)^{2/3}/ 2\, m_N \approx 1\times Y_N^{2/3}\mbox{ MeV} \ll T,}
where $n_B=N_B/V$ is the baryon number density and $Y_N=N_N/N_B$ 
with $N=n$ or $p$ the 
nucleon number fraction.
We see that neutrons and protons are nondegenerate. 
In the case of neutron stars, we
have $T=100\mbox{ keV}$ and 
$\rho_B \approx 2 \rho\subs{nuc} \approx 5.6 \times 10^{14}
\mbox{ g cm}^{-3}$,
where $\rho\subs{nuc} = 2.8 \times \times 10^{14}
\mbox{ g cm}^{-3}$ is the nuclear density.
In Appendix~\ref{NSAppendix} we will show that the proton number
fraction $Y_p$ is around 0.01 so that
\fn{\varepsilon_{F,n}\approx 94\mbox{ MeV} \gg T ,\qquad 
\varepsilon_{F,p}\approx 4\mbox{ MeV} \gg T ,}
i.e.\ both neutrons and protons are degenerate.
\subsection{Nondegenerate Limit\label{NDLaxEm}}
For a nondegenerate nucleon gas the phase space integration
(\ref{PhaseSpace}) can not be done analytically.
To simplify the squared
matrix elements (\ref{matrixcomplete}) and (\ref{matrixcompleteNP}), 
we first neglect the pion mass $m_\pi$ in the pion propagator. 
One then obtains
\fn{\sum_s |{\cal M}|_{NN}^2 = 
\frac{256 \pi^2 \alpha_\pi^2}{3 m_N^2}\,\tilde{g}_{NN}^2 \label{matrixelement}}
with
$$\tilde{g}_{NN}^2 \equiv \left\{
\begin{array}{ll}
\rule[-4mm]{0cm}{0.5cm}g_{an}^2\,(3 - \beta)&\quad n n\rightarrow n n +a\\
 \rule[-4mm]{0cm}{0.5cm}g_{ap}^2\,(3 - \beta) &\quad p p \rightarrow p p +a\\ 
{\displaystyle\left(\frac{\displaystyle g_{an} + g_{ap}}{\displaystyle2}
\right)^2(2- 4 \beta/3) 
+  \left(\frac{\displaystyle g_{an}^2 + g_{ap}^2}{\displaystyle2}\right)
(5-2 \beta /3)} &\quad n p \rightarrow n p +a\,.
\end{array} \right.$$
Here, $\tilde{g}_{NN} = \tilde{C}_N m_N/f_a$ is an effective coupling and
$\beta \equiv 3 \langle ({\bf\hat{k} \cdot \hat{l}})^2 \rangle$. In the
nondegenerate limit it turns out that $\beta=1.3078$ \cite{raffeltseckel}.
Numerically, one finds with the axion-nucleon couplings
of expression (\ref{gapganNum})
\fn{|\tilde{C}_N| = \left\{
\begin{array}{ll}
0.013 &\quad n n \rightarrow n n +a\\
0.442 &\quad p p \rightarrow p p +a\\ 
0.495 &\quad n p \rightarrow n p +a\,.
\end{array}\right.\label{totiiee}}

However, neglecting the pion mass will not be
appropriate if one considers accretion discs: in a nondegenerate
thermal medium of
nonrelativistic nucleons with temperature $T$, the 
momenta of the nucleons are approximately
given through ${\bf p}_i^2 \approx 3 m_N T$. Since the temperatures in
accretion discs are $T \approx 10\mbox{ MeV}$ or below, typical
nucleon momenta are $|{\bf p}_i| \approx 170\mbox{ MeV}$, comparable to
the pion mass $m_\pi = 135\mbox{ MeV}$. Hence, in accretion discs one
expects the pion masses to affect the axion emission rates so that we
introduce the ``fudge factor'' $\xi(T)$ which 
includes all pion-mass effects. 
A rough estimate of $\xi(T)$ in Appendix
\ref{AppPionMass} yields approximately 0.5.
With a constant matrix element
it is then straightforward to perform the integration in expression
(\ref{PhaseSpace}). One obtains
for the total energy loss rate per
unit volume
\begin{eqnarray}
Q_a\sups{ND} &=&\frac{\xi(T)}{280}\frac{T^{7/2}\, n_B^2}{m_N^{5/2} 
\pi^{7/2}}
\left(Y_n^2\,\sum_s |{\cal M}|_{nn}^2 +
Y_p^2\,\sum_s |{\cal M}|_{pp}^2 +
4\,Y_n Y_p \sum_s |{\cal M}|_{np}^2 \right)\nonumber\\
&=&
\frac{32}{105}\xi(T)
\frac{\alpha_\pi^2\, T^{7/2}\, n_B^2}{m_N^{9/2} \pi^{3/2}}
\left(Y_n^2\, \tilde{g}_{nn}^2 +Y_p^2\, \tilde{g}_{pp}^2+
4 \,Y_n Y_p\, \tilde{g}_{np}^2\right)\nonumber\\
&\equiv&
\frac{32}{105}\xi(T)
\frac{\alpha_\pi^2\, T^{7/2}\, n_B^2}{m_N^{9/2} \pi^{3/2}}
g\subs{ND}^2\, ,
\end{eqnarray}
where $g\subs{ND}$ is the total effective axion-nucleon coupling 
constant for the nondegenerate limit.
With $Y_p\approx0.1$, $Y_n\approx0.9$, and (\ref{totiiee}) we find numerically
\fn{g\subs{ND}= 4.71 \times 10^{-8}\;m_a \qquad \mbox{or}
\qquad C_N\sups{ND}=0.30\,.}
Finally, we arrive at an
axion energy loss rate per unit volume of
\fn{Q_a\sups{ND} = 4.89 \times 10^{27}\;\mbox{erg}\, \mbox{cm}^{-3}
\mbox{s}^{-1}\, T\subs{MeV}^{3.5}\,
 \rho_{12}^2\,m\subs{eV}^2 }
and at
\fn{L_a\sups{ND} = 9.7 \times 10^{48}\;\mbox{erg}\;\mbox{s}^{-1}\,
T\subs{MeV}^{3.5}\,
 \rho_{12}\,\left(\frac{M}{M_\odot}\right)m\subs{eV}^2 .
\label{axlumnondegAD}}
for the total axion luminosity.
$M_\odot$ is the solar mass and $M$ the one of the accretion torus.
\subsection{Degenerate Limit}
In the degenerate limit, the parameter $\beta$ is zero so that
we obtain from (\ref{matrixelement})
the  squared matrix elements
\fn{\sum_s |{\cal M}|_{NN}^2 = 
\frac{256 \pi^2 \alpha_\pi^2}{m_N^2}\,\tilde{g}_{NN}^2 \label{matrixDLim}}
with
$$\tilde{g}_{NN}^2 = \left\{
\begin{array}{ll}
g_{an}^2 &\quad n n \rightarrow n n +a\\
g_{ap}^2 &\quad p p \rightarrow p p +a\\ 
g_{an}^2 + g_{ap}^2 + g_{ap}^2 g_{ap}^2 /3 
&\quad n p \rightarrow n p +a\,.
\end{array}\right.$$
Numerically, the effective coupling is
\fn{|\tilde{C}_N| = \left\{
\begin{array}{ll}
0.01 &\quad n n \rightarrow n n +a\\
0.34 &\quad p p \rightarrow p p +a\\ 
0.338 &\quad n p \rightarrow n p +a\,.
\end{array}\right.\label{totiieeee}}
With these matrix elements one can determine the axion
volume emission rate. In the degenerate limit, there is no
need for introducing a fudge factor $\xi(T)$ because 
the integration in (\ref{PhaseSpace})
can be performed analytically without
neglecting the pion masses. It turns out
that the $m_\pi=0$ rates must be supplemented with a factor~\cite{RaffeltSaLfFP}
\begin{eqnarray}
F(u) = 1 - \frac{5 u}{6} \arctan\left(\frac{2}{u}\right) + 
\frac{u^2}{3(u^2+4)}
+\frac{u^2}{6\sqrt{2u^2 +4}}\nonumber\\
\times  \arctan\left(\frac{2 \sqrt{2u^2 +4}}
{u^2}\right) ,
\end{eqnarray}
where $u\equiv m_\pi / p_{F,N}$. For a \ns with 
$\rho_B \approx 2 \rho\subs{nuc}$, one has
$u \approx 0.32\,Y_N^{-1/3}$ so that we may write $F(Y_N)$.
Finally, we find for the total emission rate \cite{RaffeltSaLfFP}
\begin{eqnarray}
Q_a\sups{D} &=&\frac{31}{967680}\left(\frac{3\, n_B}{\pi}\right)^{1/3}
T^6\left(Y_n^{1/3}\,F(Y_n)\sum_s |{\cal M}|_{nn}^2+
Y_p^{1/3}\,F(Y_p)\sum_s |{\cal M}|_{pp}^2 \right.\nonumber \\ & &
\hspace{7.4cm}+
\left.4\, Y_{np}^{1/3}\,F(Y_{np})\sum_s |{\cal M}|_{np}^2\right)
\nonumber \\
&=&\frac{31\pi^{5/3}(3 n_B)^{1/3} \alpha_\pi^2\, T^6}
{3780\, m_N^2}\left(Y_n^{1/3}\,F(Y_n)\,\tilde{g}_{an}^2 +
Y_p^{1/3}\,F(Y_p)\,\tilde{g}_{ap}^2 +Y_{np}^{1/3}\,F(Y_{np})\,\tilde{g}_{np}^2
\right)\nonumber\\
&\equiv& \frac{31\pi^{5/3}(3 n_B)^{1/3} \alpha_\pi^2\, T^6}
{3780\, m_N^2}\,g\subs{D}^2\,,
\label{brinkiturn}
\end{eqnarray}
where 
$g\subs{D}$ is the total effective axion-nucleon coupling constant for
the degenerate limit and $Y_{np}$ the effective 
nucleon fraction for the mixed processes.
According to Brinkmann and Turner \cite{brinkmann}, the latter
is given by
\fn{Y_{np}^{1/3}=\frac{1}{2\sqrt{2}}\left(Y_n^{2/3}+Y_p^{2/3}\right)^{1/2}
\left[2 -\frac{|Y_n^{2/3} - Y_p^{2/3}|}{Y_n^{2/3} + Y_p^{2/3}}\right] .}
With $Y_p = 0.01$ and $Y_n = 0.99$ one has
$Y_{np}=0.06$, and we can determine the suppression due to non-vanishing
pion masses for the $nn$-, $pp$-, and $np$-processes,
$F(Y_n) \approx 0.64$, $F(Y_p)\approx 0.12$, and $F(Y_{np})\approx
0.31$. Altogether, we arrive at the following
effective coupling constant
\fn{g\subs{D}= 2.04 \times 10^{-8}\;m_a \qquad \mbox{or}
\qquad C_N\sups{D}=0.13\,.\label{EffgaNNS}}
This together with (\ref{brinkiturn}) leads to the following
expressions for the axion emission rate per unit volume,
\fn{Q_a\sups{D} = 3.74 \times 10^{29}\mbox{ erg cm}^{-3}\mbox{ sec}^{-1}\,\,
T\subs{MeV}^6\,m\subs{eV}^2\,
\left(\frac{\rho\subs{NS}}{\rho\subs{nuc}}\right)^{1/3}\;,}
where $m\subs{eV}\equiv m_a / \mbox{eV}$ and
$T\subs{MeV} \equiv T / \mbox{MeV}$.
Finally, we calculate the total axion luminosity $L_a\sups{D}$ and obtain 
\fn{L_a\sups{D} = 3.4 \times 10^{49}\mbox{ erg yr}^{-1}\,m\subs{eV}^2\,
\left(\frac{M}{M_\odot}\right)
\left(\frac{\rho\subs{NS}}{\rho\subs{nuc}}\right)^{-2/3}\,T_9^6\ ,
\label{LumDeg}}
where $M$ is
the mass of the \ns and
$T_9\equiv T/10^9\,\mbox{K}$.
\section{Axion Absorption in a Nuclear Medium\label{newsacson}}
The energy loss of a star due to particle emission depends on
whether these particles are trapped or not.
In this section we will show that axions stream away freely both
from \nss and accretion discs.
 
The dominant absorption process for axions is inverse
axion bremsstrahlung $N + N + a\rightarrow N + N$.
The associated mean free path $\lambda_a$ of the axion
is related to its absorption rate 
$\Gamma\subs{abs}$
through $\lambda_a^{-1} = \Gamma\subs{abs}$. 
One obtains
\begin{eqnarray}
\lambda_a^{-1} = \Gamma_{\mbox{\scriptsize abs}} &=&e^{E_a/T}\, 
\int \prod_{j=1}^{4}\frac{d^3 p_j}{2 E_j
(2 \pi)^3} \left(\frac{1}{2 E_a}  \right)
\,S 
\sum_s |{\cal M}|^2
\, f_1 f_2 (1-f_3)(1-f_4)\nonumber \\& &
 \times\ (2 \pi)^4\, \delta^{(4)}(P_1 + P_2 - P_3 - P_4 - P_a),
\label{schoenerTag}
\end{eqnarray}
where $E_a$ is the axion energy.
The factor $e^{E_a/T}$ accounts for the detailed-balance relationship which
reveals that
$\Gamma\subs{em}= \exp(-E_a/T)\,\Gamma\subs{abs}$,
where $\Gamma\subs{em}$ is the axion emission rate.
A comparison between (\ref{PhaseSpace}) 
and (\ref{schoenerTag}) shows that the volume emission
rate $Q_a$ is related to the mean free path by
\fn{Q_a = \frac{1}{(2\pi)^3}\int_0^\infty \lambda_a^{-1} e^{-E_a/T} 
E_a\, dp_a^3\, .}

In the case of the nondegenerate limit, one 
obtains for the axion's inverse mean
free path~\cite{RaffeltSaLfFP}
\fn{
{\lambda_a\sups{ND}}^{-1} = 
\frac{\xi(T)\pi^{1/2} \alpha_\pi^2\,n_B^2}
{6\,m_N^{9/2}\,T^{1/2}}\,
\frac{\sqrt{1+ x \pi / 4}}{x}\,g\subs{ND}^2\,,}
where $x\equiv E_a / T$. 
For freely streaming axions, the average energy is
determined by the nondegenerate nucleon-nucleon bremsstrahlung axion 
spectrum. We find 
$\langle x \rangle = \langle E_a \rangle /T = 16/7$ so that
\fn{
{\lambda_a\sups{ND}}^{-1} =  3.09 \times 10^{-9}\;
\mbox{cm}^{-1}\,T_{\mbox{\scriptsize MeV}}^{-0.5}\,\rho_{12}^2\, 
m_{\mbox{\scriptsize eV}}^2}
or
\fn{\lambda_a\sups{ND} \approx 3240\mbox{ km}\;
T_{\mbox{\scriptsize MeV}}^{0.5}\,
\rho_{12}^{-2}\, m_{\mbox{\scriptsize eV}}^{-2}.
\label{WeglangeNonDeg}}
In accretion discs, the maximum temperatures and densities are
10~MeV and $10^{12}\mbox{ g cm}^{-3}$, respectively. Then, one obtains
for the mean free path of an 20~eV axion $\lambda_a \approx 26\mbox{ km}$.
The maximum geometrical dimensions of accretion discs are about 50~km
so that it is not clear if axions escape freely. However, as we have used
extreme vales for $T$ and $\rho$, the average mean free path will be
larger than 26~km so that freely streaming axions are a good approximation.  

In the degenerate case, the inverse mean free path is 
given by \cite{RaffeltSaLfFP}
\fn{{\lambda_a\sups{D}}^{-1} = 
\frac{(3 n_B)^{1/3} \alpha_\pi^2\,T^2}{24 \pi^{7/3} m_N^2}
\frac{(x^2 + 4 \pi^2)}{(1-e^{-x})}\,g\subs{D}^2}
or numerically,
\fn{\lambda_a\sups{D} \approx 8350\mbox{ km}\;m\subs{eV}^{-2}\,
T_9^{-2}\,\left(\frac{\rho\subs{NS}}{\rho\subs{nuc}}\right)^{-1/3} ,
\label{mfpdegenerate}}
where we have used the average axion energy $\langle x \rangle = 
\langle E_a \rangle /T \approx 3.16$, which can be calculated using the
degenerate nucleon-nucleon bremsstrahlung axion 
spectrum \cite{RaffeltSaLfFP}.
For a neutron star with $T\approx 10^9\mbox{ K}$ and 
$\rho \approx 2 \rho\subs{nuc}$ we obtain
\fn{\lambda_a\sups{D} \approx 6630\mbox{ km}\;
m\subs{eV}^{-2}\,.\label{axifnpdeg}}
Axions of mass $m_a$ are not trapped inside a \ns of radius
$R$ if $\lambda_a \simgt R$.
For $R \approx 10\mbox{ km}$ this is fulfilled for
\fn{m_a \simlt 26\mbox{ eV}\,.}
Therefore, axions in the hadronic axion window are allowed to stream 
freely from \ns interiors.
As we will
discuss more precisely in Chapter \ref{sectBounds}, one expects parts of the 
neutron-star matter to be in a superfluid state. If so, the
mean free path (\ref{axifnpdeg}) is significantly enhanced.

\setcounter{equation}{0} \chapter{Impact on Black Hole Accretion Discs and
Neutron Stars\label{novelimpacts}}
We explore how the emission of axions with parameters in the hadronic
axion window affects the evolution of gamma-ray bursts.
We assume that a subclass of short bursts with a duration
of 0.1--1 seconds is driven by the accretion of hot plasma by
a black hole. Axion production in this accretion torus would 
provide a significant energy loss, altering the final $\gamma$-fluence.
Moreover, we investigate the impact of an hadronic-axion energy loss
on the cooling timescale of neutron stars. We find a relation between
their present surface temperature and the axion mass.
Our results are then compared with the observational data of the pulsars 
PSR 1055-52, PSR 0630+178 (Gemina), and PSR 0656+14.
Finally, we discuss effects of nucleon superfluidity on our
calculation.
\section{Gamma-Ray Bursts}
\subsection{General Picture}
Gamma-ray bursts are short and intense bursts of photons with energies in
the range between approximately \mbox{100 keV} and \mbox{1 MeV}. In fact,
gamma-ray bursts are the electromagnetically most luminous objects observed 
in the universe. Since their discovery in the late sixties,
several satellites have observed them, and
numerous theories have been developed to explain 
their physical nature. 
A milestone in the exploration of gamma-ray bursts was
a remarkable finding of the BATSE detector, which was 
launched in the spring of of 1991. BATSE measured a perfectly isotropic 
angular distribution of gamma-ray bursts and a deficiency of 
faint bursts \cite{GBRisotropy}. This led to the suggestion that
gamma-ray bursts are cosmological, an assumption that was confirmed by the
discovery of
x-ray transient counterparts by Beppo-SAX \cite{costaGRB}, and later by 
the discovery of optical
\cite{opticaltran} and radio counterparts \cite{radiotra}.
For the first time, it was possible to rule out a wide class of 
gamma-ray burst theories!

In spite of these discoveries, the origin of gamma-ray bursts remains 
unclear. However,
there exists a promising generic model which is in agreement with 
observation: the ``fireball model,'' where an ultra-relativistic
energy flow is converted into radiation. The initial energy flow can be
an electromagnetic Pointing flux or kinetic energy of highly-relativistic
particles.
As the conversion in radiation takes place in an optically thin region,
gamma-ray burst spectra are nonthermal, in agreement with
observation. 
\subsection{The BHAD Model}
The nature of the central engine that drives
the gamma-ray burst remains unclear because the ``inner engine'' 
that powers the relativistic fireball is hidden
from direct observations. While many models have been proposed, those 
currently favored are all based on the rapid accretion of matter by a 
black hole. These models are referred to as
black hole accretion disk (BHAD) models. It must be stressed that 
BHAD models are only able to explain short gamma-ray bursts with durations
in the range 0.1--1~s.
Several promising candidates for the progenitor system of a BHAD 
have been proposed,
the most popular being neutron star binaries, which we presently 
focus on.

The merger of two neutron stars leads to a hot and dense
central object with masses around $3\; M_\odot$, surrounded
by a toroidal cloud of hot plasma gas with a mass of about
$0.1$--$0.2\;M_\odot$. Within a timescale of a few milliseconds,
the $3\;M_\odot$ core will most likely collapse to a black hole.
However, a significant amount of mass, typically 0.03--0.3 $M_\odot$,
has too much angular momentum to be swallowed immediately by the black 
hole, and an \ad forms. Such \ads have masses between several
$10^{-2}\;M_\odot$ and a few $10^{-1}\; M_\odot$. Their densities are
around $10^{10}$--$10^{12}\mbox{ g cm}^{-3}$, their temperatures
in the range between 3 and $10\mbox{ MeV}$.

Within this accretion torus neutrinos of all flavors are 
produced by different processes. The bulk of the neutrino
luminosity is provided by the electron neutrinos $\nu_e$ and
$\bar{\nu}_e$, which are mainly produced by the charged-current processes
of electron and positron capture on protons and neutrons \cite{RuffertItoIII},
\begin{eqnarray}
e^- + p &\longrightarrow& n + \nu_e\ ,\nonumber\\
e^+ + n &\longrightarrow& p + \bar{\nu}_e\ .
\end{eqnarray}
Second is the muon and tau neutrino production, principally via
electron-positron pair annihilation 
\fn{e^- + e^+\longrightarrow \nu_i + \bar{\nu}_i\ ,\qquad i=\mu\mbox{ or }
\tau\, .}
Depending on the specific BHAD model, electron neutrinos are emitted
from a neutrinosphere with an outer radius of 70--100\mbox{ km}.
Moreover,
muon and tau neutrinos may be trapped, depending
on the thermodynamic properties of the torus \cite{JankaNEU}.
The total neutrino luminosities $L_\nu$ are typically in the ballpark of
$10^{53}\mbox{ erg s}^{-1}$. 

Now we come to the crucial point concerning gamma-ray bursts.
The emitted neutrinos and antineutrinos interact with each other
and annihilate into electron-positron pairs.
The efficiency of this energy deposition $E_{\nu\bar{\nu}}$
in form of an electron-positron plasma is usually expressed 
in terms of the parameter
$q_{\nu\bar{\nu}} \equiv \dot{E}_{\nu\bar{\nu}} / L_\nu$, where
$\dot{E}_{\nu\bar{\nu}}$ is the rate of energy deposition  by
neutrino-antineutrino annihilation.  
The efficiency $q_{\nu\bar{\nu}}$
depends on the
number densities and energies of the initial neutrinos.
According to \cite{JankaNEU},
typical values for $q_{\nu\bar{\nu}}$ are
around 0.004, i.e.\ only a half percent of the
initial neutrino energy is transformed into
thermal energy of an electron-positron plasma, the fireball. 
Integrated over the accretion timescale 
$\tau \approx 5$--\mbox{50 ms}, 
the total
energy deposition is of order
$10^{49}\mbox{ erg}$. Moreover, it is possible that the fireball does
not expand isotropically, but is beamed into axial jets with a fraction
$f_\Omega = 2 \delta\Omega /4 \pi$ of the whole sky.
Including a beaming factor
$f_\Omega \approx 0.1$--$0.01$, it follows that
the BHAD scenario is able to account for short
gamma-ray bursts with energies in the range of 
$10^{50}$--$10^{51} \mbox{ erg}$.
\subsection{Influence of Hadronic Axion Emission}
Let us now examine to what extent 
hadronic axions could alter the BHAD model for gamma-ray bursts.
During the process of accretion, axions could
be produced by nucleon-nucleon axion
bremsstrahlung. Clearly, in order to maintain
the BHAD model, the total axion luminosity $L_a$
may not exceed the neutrino luminosity 
$L_\nu$, which is of about $10^{53}\mbox{ erg s}^{-1}$.

We consider two BHAD models that have
been investigated in great detail \cite{JankaNEU}.
The first describes the attractive force of the central 
black hole by a Newtonian gravitational potential, which is known to
be singular at the origin $r=0$. The second 
takes general-relativistic effects into
account by allowing one to 
reproduce the existence
of a last stable circular orbit at a radius of
$3 r_s = 6 G M\subs{BH} / c^2$, where $r_s$ is the Schwarzschild radius. 
This model is referred to as the Paczy\'{n}ski-Wiita model.
The resulting temperatures $T$, densities
$\rho$, and \ad masses $M$ for both models are shown in
Tab.\ \ref{tabelleAD} as well as the mean values of these quantities.
One sees that the torus masses in the Paczy\'{n}ski-Wiita model are
much smaller than those of the Newtonian model.
This is a result of the very strong gravitational potential of
the Paczy\'{n}ski-Wiita model, implying that the accretion rates,
i.e.\ the mass accretion into the black hole per unit time, of
this model are much greater than those in the Newtonian description.
Therefore, the torus mass of the latter is significantly higher
than the one of the former model. However, it might be that the
Paczy\'{n}ski-Wiita model underestimates the torus mass.
\begin{table}[tb]
\caption{
Temperature $T$, density $\rho$, mass $M$ and neutrino
luminosity $L_\nu$ of an \ad
for the Newtonian
and the Paczy\'{n}ski-Wiita model. $\bbar{T}$, $\bbar{\rho}$, $\bbar{M}$, and 
$ \bbar{L}_\nu$ 
are the associated averaged quantities.
\label{tabelleAD}}
\begin{center}
\begin{tabular}{l c c}\hline\hline
&Newtonian model& Paczy\'{n}ski-Wiita model \\ \hline
$T\; [\mbox{MeV}]$ & 2--10& 2--10 \\
$\bbar{T}$&6&4\\
$\rho \; [10^{12}\,\mbox{g cm}^{-3}]$
&$0.01$--$1$&$0.01$--$0.3$\\ 
$\bbar{\rho}$
 &$0.3$&$0.1$\\
$M \; [M_\odot]$ &0.16--0.29&0.002--0.031\\ 
$\bbar{M}$ &0.23&0.02\\ 
$L_\nu\;[10^{52}\,\mbox{erg s}^{-1}]$
&1--12&
0.14--6.7\\
$\bbar{L}_\nu$
\rule[-2.7mm]{0mm}{8mm}&5.6 &4.4\\ \hline
\end{tabular}
\end{center}
\end{table}

\begin{figure}[tb]
\unitlength1mm
\begin{picture}(100,65)
\put(5,0){\psfig{file=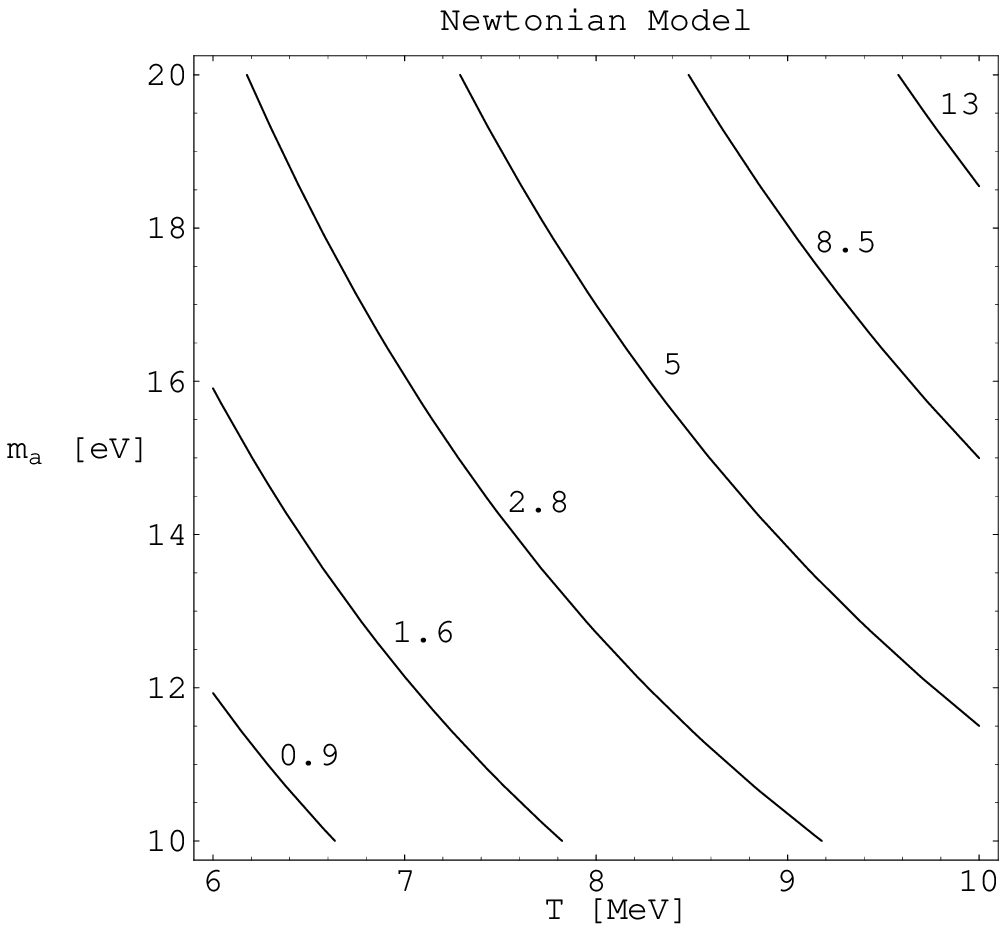,width=6cm}}
\put(70,0){\psfig{file=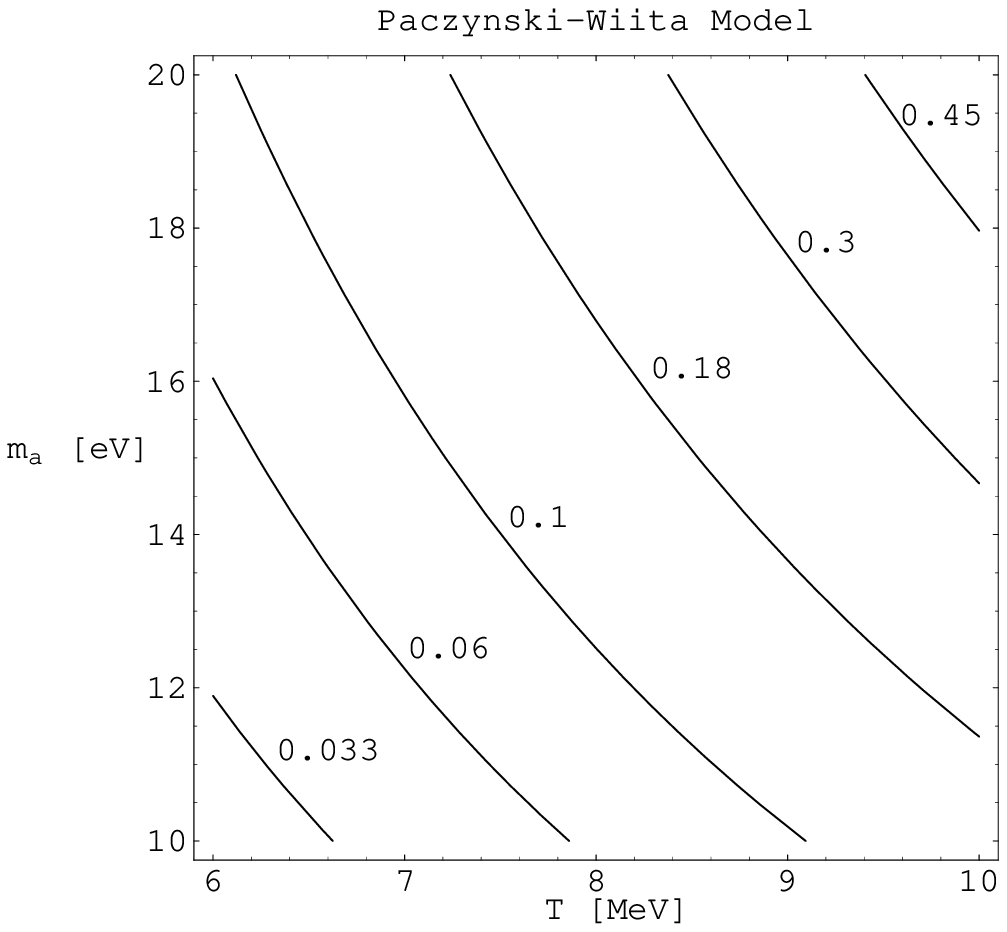,width=6cm}}
\end{picture}\par
\caption{Axion luminosity relative to the one of neutrinos 
$L_a /\bbar{L}_\nu$
as a function of the temperature $T$
and the axion mass $m_a$. The left and right panels are based on the
Newtonian and Paczy\'{n}ski-Wiita model, respectively.
\label{lastplot}}
\end{figure}

Now, we want to estimate the
axion luminosity $L_a$. As we have seen in the previous chapter, axions
stream away freely from accretion discs, implying that 
axion volume emission takes place. The corresponding
energy loss is given by expression (\ref{axlumnondegAD}).
It is linear in the mass $M$ and the density $\rho$ so that one may use
the average values given in
Tab.\ \ref{tabelleAD}. Then, one finds for the two models
\fn{L_a = \left\{
\begin{array}{cl}
6.7\times 10^{47}\;\mbox{erg}\;\mbox{s}^{-1}\,T\subs{MeV}^{3.5}\,
m\subs{eV}^2 &\quad \mbox{Newtonian model}\\
1.9\times 10^{46}\;\mbox{erg}\;\mbox{s}^{-1}\,T\subs{MeV}^{3.5}\,
m\subs{eV}^2 &\quad \mbox{Paczy\'{n}ski-Wiita model.}
\end{array} \right.\label{NewtonianAxLum}}
Now, these luminosities
have to be compared with those of the neutrinos given in Tab.\ \ref{tabelleAD}.
This is graphically done in Fig.\ \ref{lastplot}, where
the ratio
$L_a /\bbar{L}_\nu$ as a function of the temperature $T$
and the axion mass $m_a$ is plotted.

Obviously, in the case of the Newtonian model, hadronic axions could alter
the existence of gamma-ray bursts, whereas for the
Paczy\'{n}ski-Wiita model the axion luminosities are at 
least one order of magnitude
smaller than the one of the neutrinos. This is a consequence of the
small torus masses in the Paczy\'{n}ski-Wiita model.

Finally, we want to estimate the maximum axion production rate which
can be obtained for the maximum temperature $T \approx 10\mbox{ MeV}$.
With (\ref{NewtonianAxLum}) we find
\fn{L_a = \left\{
\begin{array}{cl}
2.1\times 10^{51}\;\mbox{erg}\;\mbox{s}^{-1}\,
m\subs{eV}^2 &\quad \mbox{Newtonian model}\\
6.0\times 10^{49}\;\mbox{erg}\;\mbox{s}^{-1}\,
m\subs{eV}^2 &\quad \mbox{Paczy\'{n}ski-Wiita model.}
\end{array} \right.}
As a result,
one obtains an upper bound on the allowed axion mass of approximately
\fn{m_a \simlt \left\{
\begin{array}{cl}
5\mbox{ eV}&\quad \mbox{Newtonian model}\\
27\mbox{ eV} &\quad \mbox{Paczy\'{n}ski-Wiita model} .
\end{array} \right.}

It is not clear which of both models gives an appropriate description
of the accretion process, or whether they are a good approximation 
of the black hole gravitational potential at all. Therefore, we estimate that
axions in the parameter range between 10 and 100\mbox{ eV} are in principle
able to alter the evolution of gamma-ray bursts significantly.
Note that this range includes the hadronic axion window 
of $10$--$20\mbox{ eV}$,
i.e.\ the existence of \has would be relevant for
gamma-ray burst physics. 

Presently, nobody knows whether the
BHAD model gives the correct description of gamma-ray bursts. However, 
if this picture
proves to be correct, at least for short gamma-ray bursts, 
one could
calculate reliable bounds on the axion mass.
On the other hand, if hadronic axions were detected
by the solar-axion search experiments \cite{moriyama, krcmar},
the BHAD model
certainly would have to include axion emission in a detailed description.
\section{Neutron Stars}
\subsection{A Simple Model for Axion Cooling}
We showed in the Sect.~\ref{newsacson}
that axions in the hadronic axion window are allowed to stream freely
from neutron star interiors,
implying an additional energy loss which
could have observable consequences which shall now be explored.
Therefore, we set up a simple model of late-time \ns cooling,
assuming that the star cools via axion volume emission. Furthermore,
we neglect all
other cooling mechanisms because axions only alter 
\ns cooling significantly if they are the dominant cooling channel. 
The corresponding axion luminosity was given in (\ref{LumDeg}),
\fn{L_a = 3.4 \times 10^{49}\mbox{ erg yr}^{-1}\,m\subs{eV}^2\,
\left(\frac{M}{M_\odot}\right)
\left(\frac{\rho\subs{NS}}{\rho\subs{nuc}}\right)^{-2/3}\,T_9^6\ .}
This result is based on the assumption that the 
nuclear medium can be treated
as an ideal, degenerate Fermi gas of neutrons and protons. However, the real 
nuclear equation of state of the superdense matter inside a 
\ns is still unknown. According to the BCS theory
of superfluidity, one expects parts of 
the neutron star to be in a superfluid state.
At lower densities $\rho \approx 10^{13}$--$10^{14}\mbox{g cm}^{-3}$
particles are expected to be in a
${^1\mbox{S}_0}$ superfluid state when the temperature falls below the
critical temperature $T\subs{cr}$. A second superfluid state ${^3\mbox{P}_2}$
occurs in more dense matter with 
$\rho \approx 10^{14}$--$10^{15}\mbox{g cm}^{-3}$, implying that neutrons in
the core of \nss should be in this superfluid state. As the
density of protons in a \ns is much smaller than the one of the neutrons, 
they are expected to be in the ${^1\mbox{S}_0}$ state. However,
superfluidity in \nss is a difficult matter, and many
of its aspects are not yet understood. Particularly controversial
is the proton ${^1\mbox{S}_0}$ pairing.
But, if superfluidity occurs,
it has different effects on the cooling of neutron stars; the
most important in the context with axion emission
is the suppression of our estimated axion-nucleon
interaction (\ref{EffgaNNS}). Moreover, strong superfluidity
reduces the heat capacity $C$ of the nucleon gas.
We include these effects by introducing an effective axion luminosity
and heat capacity,
\fn{L_a\sups{eff} = \frac{1}{f\subs{em}}\,L_a
\equiv \frac{1}{f\subs{em}} A\,T_9^6\qquad \mbox{and}\qquad C\subs{eff} = \frac{1}{f\subs{C}}\,C ,\label{neudefeq}}
respectively, where $f\subs{em}$ and $f_C$ are fudge factors.

The effective axion luminosity is linked with an energy loss of the
neutron star.
According to our simple model of nuclear matter, the total,
unsuppressed heat capacity
$C$ of a \ns is that
of a degenerate ideal gas of neutrons, protons, and
electrons,
\fn{
C = \frac{\pi^2}{2}T\left(
\frac{N_n}{\varepsilon_{F,n}}+\frac{N_p}{\varepsilon_{F,p}}+
\frac{N_e}{\varepsilon_{F,e}}\right) ,}
where $N_n$, $N_p$, and $N_e$ are the numbers of neutrons, protons, and
electrons and $\varepsilon_{F,n}$, $\varepsilon_{F,p}$, and
$\varepsilon_{F,ne}$ are the related Fermi energies. For nonrelativistic
nucleons $i=n, p$ one has 
$p_{F,i}= (3\pi^2 Y_i\, n_B)^{1/3}\equiv p_{F,N} Y_i^{1/3}$
and
$\varepsilon_{F,i} = p_{F,i}^2 /2 m_N \equiv \varepsilon_{F,N} Y_i^{2/3}$
whereas $\varepsilon_{F,e} = p_{F,e}$ for the relativistic electrons.  
Charge neutrality implies $n_e=n_p$ and thus $p_{F,e}=p_{F,p}$ so that
$\varepsilon_{F,e}=p_{F,p}$.
Finally we obtain
\begin{eqnarray}
C &=&\frac{\pi^2}{2}\frac{N_B T}{\varepsilon_{F,N}} \left(
Y_n^{1/3} +Y_p^{1/3}+ Y_p^{2/3}\,\frac{\varepsilon_{F,N}}{p_{F,N}}\right)
\nonumber\\
& = &\frac{\pi^2}{2}\frac{N_B T}{\varepsilon_{F,N}} \left(
0.997 + 0.215 +0.009\right) ,
\end{eqnarray}
where we have used $Y_p = 0.01$.
Obviously, neutrons, protons, and electrons contribute
around \mbox{82\%}, \mbox{17\%} and \mbox{1\%}, respectively,
to the total heat capacity $C$.
If we express $C$ in terms of the solar mass and the
nuclear density, we
find explicitly
\fn{C = 1.42\times 10^{48}\frac{\mbox{erg}}{10^9\mbox{ K}}\;T_9\,
\left(\frac{M}{M_\odot}\right)
\left(\frac{\rho\subs{NS}}{\rho\subs{nuc}}\right)^{-2/3}
\equiv B\, T_9 .\label{heatcapa}}
As axions are considered to be the dominant cooling channel, one obtains as a
result of energy conservation 
\fn{\frac{dE\subs{NS}}{dt}= C\subs{eff}\, \frac{dT}{dt}= - L_a\sups{eff} ,}
where $E\subs{NS}$ is the total thermal energy of the neutron star.
With (\ref{neudefeq}) and (\ref{heatcapa}) we obtain the equation
\fn{\frac{1}{f_C} B \, T_9\,\frac{dT_9}{dt} = - \frac{1}{f\subs{em}}
A\,T_9^6\ .}
This equation can be easily integrated with the result
\fn{\Delta t = t_f - t_i = \frac{B f\subs{sup}}{4 A}\,T_9(f)^{-4}
\left(1-\frac{T_{9}(f)^4}{T_{9}(i)^4}\right) ,\label{zwischenerg}}
where $T_9(i)\equiv T_i /10^9\mbox{ K}$ and $T_9(f)\equiv T_f /10^9\mbox{ K}$ 
are the internal temperatures of the
neutron star, taken at the initial and final
times $t_i$ and $t_f$, respectively. Furthermore, we have introduced
the total fudge factor $f\subs{sup} \equiv f\subs{em} / f_C$.
Without introducing
any large errors, 
$T_9(f)/T_9(i)$ can be taken to be zero. 
Then (\ref{zwischenerg}) simplifies to
\fn{t\subs{yr} \approx 0.010\; m\subs{eV}^{-2}\,{f\subs{sup}}\,
T_9(f)^{-4}}
or
\fn{T_{C9} = 0.32\;f\subs{sup}^{1/4}\, m\subs{eV}^{-1/2}\,
t\subs{yr}^{-1/4}\ ,
\label{Bezeins}}
where $t\subs{yr}\equiv t_f / \mbox{yr}$ is the \ns age in years 
and $T_{C9} \equiv T_9(f)$.
It should be stressed that $T_{C9}$ is effectively the core temperature
because we do not consider axion production in the \ns crust.
Consequently, equation (\ref{Bezeins}) allows us to calculate the 
internal temperature $T_C$ of a 
\ns depending on its age $t\subs{yr}$, the axion mass $m_a$, and
the fudge factor ${f\subs{sup}}$.
A remarkable feature of the
solution (\ref{Bezeins}) is that it depends neither on
the mass $M$ nor on the density $\rho$ of the neutron star.

\begin{figure}[ttt]
\unitlength1mm
\begin{picture}(70,70)
\put(8,-5){\psfig{file=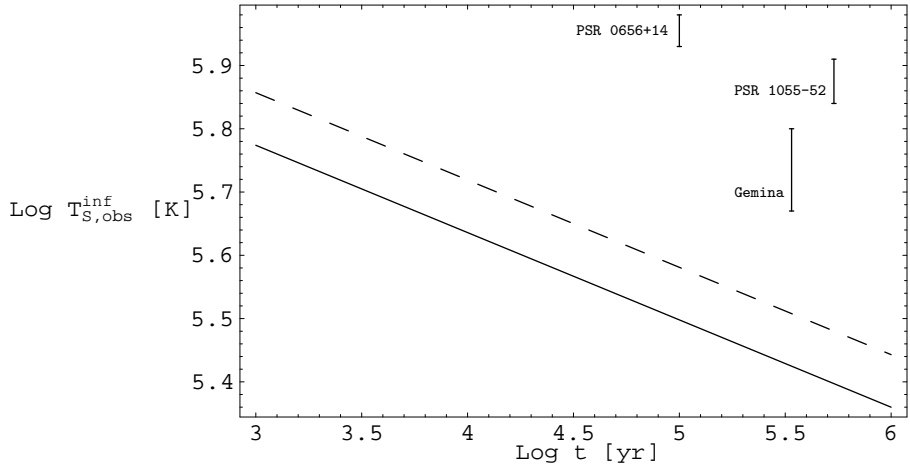,width=12.0cm}}
\put(105,51){\tiny \tt PSR 1055-52}
\put(105,37.5){\tiny \tt Gemina}
\put(84,59){\tiny \tt PSR 0656+14}
\end{picture}\par
\caption{Red-shifted surface temperature 
$T_{S, \mbox{\scriptsize obs}}^\infty$ versus age for axion
cooling. The dashed and solid curves correspond to
axion masses of \mbox{10 eV} and \mbox{20 eV}, respectively.
The fudge factor ${f\subs{sup}}$ was assumed to be $10^3$.} 
\label{axioncoolplot}
\end{figure}

In order to compare our prediction with observational data, it
is appropriate to rewrite equation (\ref{Bezeins}) in terms of the
observable
surface temperature $T_{S}$, instead of the internal temperature $T_{C}$.
Strictly speaking, the temperature detected by a distant observer is not
$T_{S}$, but the red-shifted temperature 
$T_S^\infty = T_S \sqrt{1-R_g/R}$, where $R_g$ is the gravitational radius.
For neutron stars, the stellar radius $R$ is only
2--3 times larger than $R_g$, i.e.\ $R/R_g = 2.5$. 
Therefore, we obtain the relation
\fn{T_{S} \approx 1.29\,T_S^\infty\ .\label{redshifttemp}} 
Furthermore,  
as a consequence of thermal conductivity, the core temperature $T_C$ 
is related to the surface temperature $T_S$.
Approximately, the following
equation holds \cite{GPIfotmula}:
\fn{T_{C9} = 0.128 \, \left(\frac{T_{S6}^4}{g_{14}}\right)^{5/11}\ ,
\label{Bezzwei}}
where $T_{S6}\equiv T_S /10^6\mbox{ K}$ and $g_{14}$ is the gravitational
acceleration on the \ns surface in units of $10^{14}\mbox{ cm sec}^{-2}$.
Eliminating the temperature $T_{C9}$ between the equations
(\ref{Bezeins}) and (\ref{Bezzwei}), leads together with (\ref{redshifttemp}) 
to
\fn{T_{S6}^\infty = 1.282 \; g_{14}^{1/4}\;{f\subs{sup}}^{11/80}\;
m\subs{eV}^{-11/40}\;
t\subs{yr}^{-11/80}\ .}
The gravitational surface acceleration is typically in the range of
$g_{14} \approx 2$--3. Therefore, with $g_{14} = 2.5$ and the logarithmic 
scales $\log T_{S}$ and $\log t\subs{yr}$, we finally arrive at
 \fn{\log T_{S}^\infty \approx 
0.138\,\log\left(\frac{{f\subs{sup}}}{m\subs{eV}^2}\right) -
0.138\, \log t\subs{yr} + 6.208 ,\label{finalTestBez}}
where $\log$ is the logarithm to base 10.
Now we are in a position to compare our predictions with the observational 
data
from ROSAT. Axions with masses $m_a$ are execluded if the predicted 
temperatures
$\log T_{S}^\infty$ of expression
(\ref{finalTestBez}) are below those observed, i.e.\ 
$T_{S}^\infty < T_{S, \mbox{\scriptsize obs}}^\infty$.
The measured effective surface temperatures
$T_{S, \mbox{\scriptsize obs}}^\infty$ and the ages of
three pulsars PSR 1055-52 \cite{ogelmann}, PSR 0630+178 
(Gemina) \cite{halpwang}, 
and PSR 0656+14 \cite{possenti} are listed
in Tab.\ \ref{TabelleDaten}. 
\begin{table}[tb]
\caption{
Measured effective surface temperatures 
$T_{S, \mbox{\scriptsize obs}}^\infty$ 
and derived bounds on the ratio
${f\subs{sup}} / m\subs{eV}^2$, where $f\subs{sup}$ is a fudge factor
which takes the suppression of axion emission due to superfluidity
into account.\label{TabelleDaten}}
\begin{center}
\begin{tabular}{l c c l}\hline\hline
Source & $\log t\subs{yr}$ & 
$\log T_{S, \mbox{\scriptsize obs}}^\infty\;[\mbox{K}]$ &
Exclusion range for ${f\subs{sup}} / m\subs{eV}^2$\\ \hline
PSR 1055-52 \rule[-2.5mm]{0mm}{8mm}&5.73& 5.84--5.91&
$\hspace{1.3cm}{f\subs{sup}} / m\subs{eV}^2 \simlt 1167$\\
Gemina \rule[-2.5mm]{0mm}{8mm}& 5.53&5.67--5.80&
$\hspace{1.3cm}{f\subs{sup}} / m\subs{eV}^2 \simlt 43$\\ 
PSR 0656+14 \rule[-2.5mm]{0mm}{8mm}& 5.00&5.93--5.98&
$\hspace{1.3cm}{f\subs{sup}} / m\subs{eV}^2 \simlt 975$\\ \hline
\end{tabular}
\end{center}
\end{table}
Additionally, in the fourth
column the exclusion ranges for ${f\subs{sup}} / m\subs{eV}^2$
are given for each source.
To rule out the whole hadronic axion window between 10 and 
\mbox{20 eV}, we obtain maximal allowed fudge factors ${f\subs{sup}}$
of order $10^5$, $4\times10^3$, and $10^5$  
for the pulsars PSR 1055-52, Gemina, and PSR 0656+14,
respectively. 
Fig.~\ref{axioncoolplot} shows the predicted cooling curve 
of a \ns according to our model. 
The axion mass is
$m_a = \mbox{10 eV}$ and $m_a = \mbox{20 eV}$ for
the dashed and solid curves, respectively.
Furthermore, the fudge factor was supposed to be ${f\subs{sup}} = 10^3$. 
In spite of this extreme suppression of the axion emission, 
the associated axion cooling predicts present-day surface temperatures
$T_{S, \mbox{\scriptsize obs}}^\infty$ which are far below the
measured data, i.e.\ the observed \nss are too hot to
have been cooled by hadronic axion emission.
\subsection{Numerical Cooling Calculations Including Axions}
Recently, Umeda et al.\ \cite{UmedoIwamoto} have investigated the 
thermal evolution of \nss
including axionic energy losses. They intended to compete with
the low mass SN 1987A limit of equation (\ref{SNexclusion}), which implies
the upper limit $m_a \simlt 0.01 \mbox{ eV}$ on the axion mass. 
Just as we did above, they assumed axion emission due to nucleon-nucleon 
bremsstrahlung.
They based their calculation on three
equations of state, a stiff, a medium, and a soft one; corresponding
to different compressibilities of the nuclear matter.
Furthermore, they took
${^3\mbox{P}_2}$ Cooper pairing of neutrons into account. 
Umeda et al.\ compared their results with the data of
the PSR 0656+14 pulsar and obtained the limits
\fn{m_a \simlt \left\{
\begin{array}{cl}
0.33\mbox{ eV} &\quad \mbox{stiff model}\\
0.08\mbox{ eV} &\quad \mbox{medium model}\\
0.06\mbox{ eV} &\quad \mbox{soft model},
\end{array} \right.\label{katze}}
which are less stringent than the one of SN 1987A .

Nevertheless, we can gain from their calculation because it has
implications for the hadronic axion window.
Umeda et al.\ assumed
that axions do not get trapped in the interior of neutron stars. 
Hence, their constraint
applies as long as the free streaming condition $\lambda_a > R$ is 
satisfied. However, we have seen in Sect.\ \ref{newsacson} that
hadronic axions stream away freely, i.e.\ the result 
of Umeda et al.\ excludes axions with parameters of the
hadronic axion window. 

Over and above that, we can use their limits
to estimate of the fudge factor $f\subs{sup}$.
With (\ref{katze}) and the condition 
${f\subs{sup}} / m\subs{eV}^2 \simgt 975$ for the PSR 0656+14 we find
\fn{f\subs{sup} \approx \left\{
\begin{array}{cl}
106 &\quad \mbox{stiff model}\\
6.2 &\quad \mbox{medium model}\\
3.5 &\quad \mbox{soft model}.
\end{array} \right.}
The fudge factor $f\subs{sup}$ is a compound parameter taking
the suppression of axion emission and heat capacity
due to superfluidity into account. In the case of the
stiff equation of state,
neutrons are assumed to be strongly superfluid, implying a significant
reduction of their heat capacity---actually it becomes smaller than
the one of the electrons so that
neutrons
effectively do not contribute to the total heat capacity $C$ of the
neutron star.
We have seen above that neutrons contribute around 80\% 
to the total heat capacity. Therefore, for the stiff model
we estimate a heat-capacity
fudge factor $f_C$ of about 5, implying $f\subs{em} \approx 530$.
In models with a softer equation of state, the reduction of
the heat capacity as a result of superfluidity is smaller, and the
fudge factor is difficult to determine. Approximately, we use
$f_C \approx 2$ and $f_C \approx 1$ for the medium and the soft model,
respectively. Altogether we have
\fn{f\subs{em} \approx \left\{
\begin{array}{cl}
530 &\quad \mbox{stiff model}\\
12 &\quad \mbox{medium model}\\
4 &\quad \mbox{soft model}
\end{array} \right.\label{fudgeem}}  
for the suppression of the axion emission as a consequence of 
neutron superfluidity.

Now, we are in a position to specify the axion trapping
inside a neutron star more precisely. 
The mean free path $\lambda_a\sups{D}$ was given in equation (\ref{axifnpdeg}),
however,
without taking superfluidity into account. The effective
mean free path is given by $\lambda_a\sups{eff} = f\subs{em}\, 
\lambda_a\sups{D}$, i.e.\
\fn{\lambda_a\sups{eff}\approx 6630\mbox{ km}\;f\subs{em}\;m\subs{eV}^{-2}\, .}
Axions get trapped if $\lambda_a\sups{eff}$ is comparable to the 
neutron-star radius $R\approx 10\mbox{ km}$. Therefore, from
(\ref{fudgeem}) we obtain the following conditions for free streaming axions,
\fn{m_a \simlt \left\{
\begin{array}{cl}
593\mbox{ eV} &\quad \mbox{stiff model}\\
89\mbox{ eV} &\quad \mbox{medium model}\\
51\mbox{ eV} &\quad \mbox{soft model}.
\end{array} \right.}
Hence, we conclude that axions do not get trapped in
the interior of \nss if their masses are less than
50--600~eV.
\section{Is the Hadronic Axion Window Closed?}
We have re-examined the possibility of an astrophysically allowed
KSVZ-type axion which has a strongly suppressed coupling to photons, 
and we have confirmed that such ``hadronic axions''
with masses between 10 and 20~eV
are not excluded by previous arguments. Axions
in this window were thermally produced in the
early universe, implying that they constitute HDM with 
$\Omega_a \approx 0.1$--0.4, in agreement with the recent work of 
Moroi and Murayama \cite{Moroietal}.
Hence, \has not only solve
the strong CP problem, but they are also
of cosmological importance.

As hadronic axions mainly couple to nucleons, we have investigated
their impact on accretion discs and isolated neutron stars.
We have found that hadronic-axion production in \ads would have 
observable consequences if these
objects really provide an explanation for short 
gamma-ray bursts as in the BHAD model. 
This model is based on coalescing compact binaries such as two neutron stars,
implying the emission of gravitational waves which could in future be
detected by the laser interferometers LIGO, VIRGO, TAMA, and GEO
\cite{frederic}. If so, a time-coincident observation
of gamma-ray bursts and gravitational waves would be strong evidence
for the BHAD model. However, even if the BHAD model proves to be true,
it is premature to rule out hadronic axions
because their emission rate from the accretion torus depends significantly
on details of the BHAD model. On the other hand, a future detection of
solar \has via resonant absorption in ${^{57}\mbox{Fe}}$
\cite{moriyama, krcmar} would have important consequences
for the BHAD scenario, perhaps even calling this
explanation for gamma-ray bursts into question.

The most significant limits on hadronic axions arise from old neutron stars.
Our simple cooling model indicates that
\has accelerate the cooling process of these objects 
significantly, with the consequence that \nss today are actually
too hot to be in accord with hadronic axions. Even in the
presence of superfluidity, which suppresses the
emission rates, this discrepancy remains significant.
The same conclusion is reached on the basis of the 
numerical neutron-star cooling simulations by Umeda 
et al.\ \cite{UmedoIwamoto}.
Conversely, if the existence of \has were verified, the standard
cooling mechanism for \nss would have a serious blemish. In this case, 
internal heating effects would have to be included \cite{Christoph} or
the interpretation of the observed soft x-ray components of the 
pulsars PSR 1055-52, Gemina, 
and PSR 0656+14 \cite{ogelmann,halpwang,possenti}
as thermal blackbody spectra would have to be reconsidered. 

\renewcommand {\theequation}{\Alph{chapter}.\arabic{equation}}
\begin{appendix}
\setcounter{equation}{0} \chapter{Pion Mass Effects in the
Bremsstrahlung Process \label{AppPionMass}}
It was mentioned in Chapter \ref{nucmedres} that the influence of non-zero pion masses $m_\pi$ on the axion emission
via nucleon-nucleon axion bremsstrahlung is a temperature dependent effect, and
that the necessary phase-space integrations
can not be done analytically in the case of the
nondegenerate limit.
It is therefore appropriate to estimate these effects as a 
function of the temperature.
Recall that the squared matrix element for 
nucleon-nucleon axion bremsstrahlung is 
\begin{eqnarray}
\sum_s |{\cal M}|^2 = \frac{256 \pi^2}{3}\frac{g_{aN}^2 \alpha_\pi^2}{m_N^2} 
\left[
\left(\frac{{\bf k}^2}{{\bf k}^2 + m_\pi^2}\right)^2  + 
\left(\frac{{\bf l}^2}{{\bf l}^2 + m_\pi^2}\right)^2  \right. \nonumber \\
 \left.  +  \frac{{\bf k}^2\, {\bf l}^2 - 3 ({\bf k \cdot l})^2}{({\bf k}^2 + m_\pi^2)({\bf l}^2 + 
m_\pi^2)}\right] ,\label{app1matr}
\end{eqnarray}
where ${\bf k} = {\bf p}_2 - {\bf p}_4$ and ${\bf l} = {\bf p}_2 - {\bf p}_3$ 
with the nucleon's momenta ${\bf p}_i,\ i=1\ldots 4$.
For zero pion masses (\ref{app1matr}) becomes
\fn{\left.\sum_s |{\cal M}|^2\right|_{m_\pi = 0} = 
\frac{256 \pi^2}{3}\frac{g_{aN}^2 \alpha_\pi^2}{m_N^2}
(3 - \beta)}
with $\beta \equiv 3 \langle ({\bf\hat{k} \cdot \hat{l}})^2 \rangle = 1.3078$.

\begin{figure}[tb]
\unitlength1mm
\begin{picture}(100,80)
\put(22,0){\psfig{file=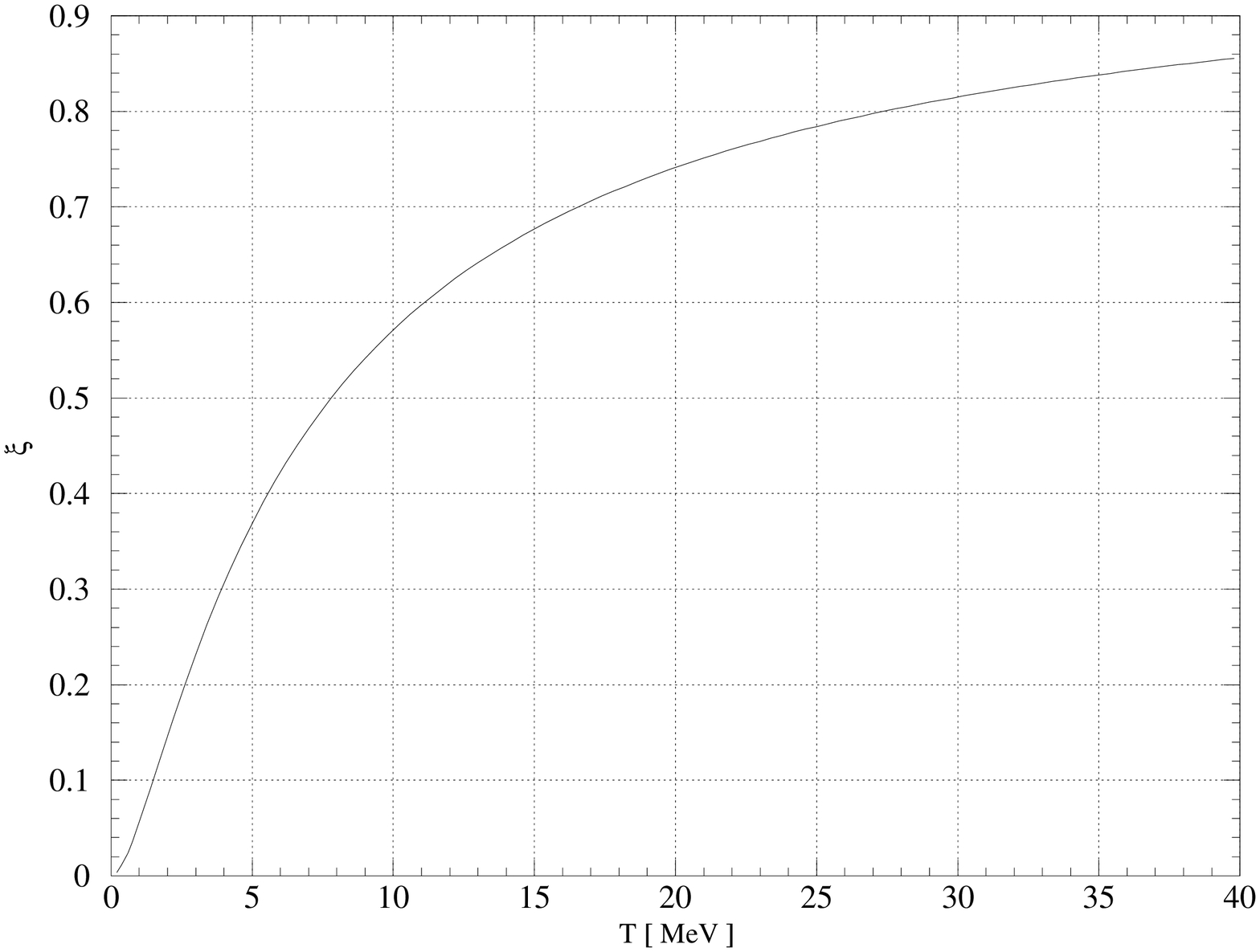,width=10.0cm}}
\end{picture}\par
\caption{The factor $\xi(T)$ as defined in (\ref{XivoTdefinition}).} 
\label{XivonTPlot}
\end{figure}

In a thermal medium the momenta
of nonrelativistic nucleons are on average
$\langle{\bf p}^2\rangle \approx 3 m_N T$, so
that one can estimate ${\bf k}^2 = {\bf l}^2 \approx 6 m_N T$. 
Then, we find for the momentum dependent terms 
of (\ref{app1matr})
\fn{\left(\frac{{\bf k}^2}{{\bf k}^2 + m_\pi^2}\right)^2 \approx \left(\frac{{\bf l}^2}{{\bf l}^2 + m_\pi^2}\right)
^2 \approx \left(\frac{1}{1+\frac{m_\pi^2}{6 m_N T}}\right)^2 \equiv \xi(T) \label{xi} \label{XivoTdefinition}}
and
\fn{\frac{{\bf k}^2\, {\bf l}^2 - 3 ({\bf k \cdot l})^2}{({\bf k}^2 + m_\pi^2)({\bf l}^2 + 
m_\pi^2)} \approx \xi(T) - \beta\,\xi(T)\ .}
Altogether, we finally obtain for the squared matrix element
\begin{eqnarray}
\sum_s |{\cal M}|^2 &=& 
\frac{256 \pi^2}{3}\frac{g_{aN}^2 \alpha_\pi^2}{m_N^2}
\xi(T)\, 
\left(3 - \beta \right)\\ &\equiv& \xi(T)\, \left.\sum_s |{\cal M}|^2\right|_{m_\pi = 0}\ ,
\end{eqnarray}
i.e.\ all effects attributed to a non vanishing pion mass $m_\pi$ are combined in the overall
factor $\xi(T)$. In Fig.\ \ref{XivonTPlot}, the function $\xi(T)$ is plotted for temperatures between
0 and \mbox{40 MeV}. 
This simple estimation of pion mass effects is in good agreement with the exact calculation of
Raffelt and Seckel \cite{raffeltseckel}. However, our approximation underestimates the reduction
of axion emission rates due to the pions mass: At temperatures around \mbox{10 MeV} the exact
result predicts a reduction of about \mbox{50\%},
in contrast to our estimation, which yields
\mbox{57\%}.
\chapter{Nucleons and Electrons in Neutron Stars\label{NSAppendix}}
We consider a nuclear medium which is made up of protons, neutrons, electrons
and neutrinos. We assume that all particle species
are in chemical equilibrium with each other
as result of the reaction $e + p \leftrightarrow n+ \nu_e$. In addition,
neutrinos are allowed to stream freely away, i.e.\ their chemical
potential vanishes $\mu_\nu =0$.
The phase-space occupation function of each species $i = p, n, e, \nu_e$
is given by the Fermi-Dirac distribution
\fn{f({\bf p}_i) = \frac{1}{\exp[(E_i({\bf p}_i) - \mu_i)/T] + 1}\ ,
\label{fermidiridistAPP}}
with the energy 
$E_i({\bf p}_i)^2 = m_i^2 + {\bf p}_i^2 $ and
the chemical potential $\mu_i$.
The value of $\mu_i$ at $T=0$ defines
the Fermi energy $E_{F,i}^2 = m_i^2 + p_{F,i}^2$ in terms of the Fermi
momentum $p_{F,i}$, which is related to the
particle's number density by $n_i =g\, p_{F,i}^3/ 6 \pi^2$ with
$g=1$ for neutrinos and $g=2$ for nucleons and electrons.
For a nonrelativistic particle species (neutrons and protons 
in neutron stars),
it is appropriate to use the nonrelativistic
Energie $E\sups{kin}_i = E_i({\bf p}_i) -m_i\approx {\bf p}_i^2 / 2\, m_i,
\ i=n,p$ and the nonrelativistic chemical potential 
$\hat{\mu}_i \equiv \mu_i -m_i$.

It is useful to introduce the neutron and proton number fractions 
$Y_n$ and $Y_p$, respectively. They are defined as
\fn{Y_n = n_n / n_B\quad \mbox{and} \quad Y_p = n_p / n_B\ ,}
where $n_B \equiv (N_p + N_n)/V$ is the total baryon number density.
Consequently, $Y_p + Y_n =1$.
Furthermore, the electron fraction $Y_e= n_e/n_B$ describes the number 
of electrons
per baryon. As a result of charge neutrality, the electron fraction is equal
to the proton fraction,
$Y_e = Y_p$.

The number fractions $Y_p$ and $Y_n$ can be determined easily
for neutron stars
if one works on the assumption that the star is
transparent to neutrinos, i.e.\ $\mu_{\nu_e} =0$. This is justified since
we cosider quite cold neutron stars.
One then obtains the proton fraction in the case of a degenerate medium
\cite{StuartSaul}
\fn{Y_p = \frac{n_p/n_n}{n_p/n_n + 1}\label{ProtFrac}}
with
\fn{\frac{n_p}{n_n} \approx \frac{1}{8}\left(\frac{
1 + 4 Q / m_n\,x_n^2 + 4 (Q^2 -m_e^2) / m_n^2\,x_n^4}{
1+1/x_n^2}\right)^{3/2},}
where $Q =m_n - m_p = 1.29\mbox{ MeV},\ x_n=p_{F,n}/m_n$, and
$p_{F,n}$ is the Fermi momentum of the 
neutron.
From that, the neutron and electron number fractions follow  
immediately through $Y_n = 1 - Y_p$ and $Y_e = Y_p$.

We are concerned with \nss which have characteristic densities
of $\rho_B \approx 2 \rho\subs{nuc} \approx 5.6\times
10^{14}\mbox{ g cm}^{-3}$ and
core temperatures less than $100\mbox{ keV}$.
Then, one explicitly finds $Y_p \approx 0.009$, i.e.\ nuclear matter in
\nss is
approximately made up of 99\% 
neutrons and 1\% 
protons. However, it should be stressed that these values depeds
strongly on the underlying equation of state. Moreover, a \ns is not
just a homogeneous ball but consists diverent layers
like the outer crust, inner crust, outer core, and inner core.
Another problem is that in a dense medium
nucleon interactions become important
so that the ideal gas is no longer a good approximation of the equation of
state. As a consequence, the nucleon effective mass $m^\ast_N$ is less
than its vacuum value, and its thermal excitations depend on
a uncertain dispersion relation.
Summing up, we conclude that the proton fraction in \ns is
a few percent of the neutron number.
\end{appendix}

\end{document}